\documentclass[11pt,a4paper]{article}
\pdfoutput=1
\usepackage{jheppub}
\usepackage{microtype}
\usepackage{rotating}
\usepackage{booktabs}
\usepackage{multirow}
\usepackage{tabularx}
\usepackage{dcolumn}
\usepackage{bm}
\usepackage[version=3]{mhchem}

\newcolumntype{R}{>{\raggedleft\arraybackslash}X}
\newcolumntype{Y}{>{\centering\arraybackslash}X}

\newcommand{\bb}{\ensuremath{\beta\beta}}
\newcommand{\bbonu}{\ensuremath{0\nu\beta\beta}}
\newcommand{\bbtnu}{\ensuremath{2\nu\beta\beta}}

\newcommand{\mbb}{\ensuremath{m_{\bb}}}
\newcommand{\Qbb}{\ensuremath{Q_{\bb}}}
\newcommand{\Xe}{\ensuremath{^{136}\mathrm{Xe}}}
\newcommand{\Bi}{\ensuremath{^{214}\mathrm{Bi}}}
\newcommand{\Tl}{\ensuremath{^{208}\mathrm{Tl}}}
\newcommand{\ckky}{\ensuremath{\mathrm{counts/(keV~kg~yr)}}}
\newcommand{\hair}{\ifmmode\mskip1mu\else\kern0.08em\fi}

\makeatletter
\g@addto@macro\bfseries{\boldmath}
\makeatother

\begin{document}

\title{Sensitivity of NEXT-100\\ to neutrinoless double beta decay}
\collaboration{The NEXT Collaboration}
\author[a,1]{J.~Mart\'in-Albo,\note{Corresponding author. Now at University of Oxford, United Kingdom.}}
\author[a]{J.~Mu\~noz Vidal,}
\author[a]{P.~Ferrario,}
\author[a]{M.~Nebot-Guinot,}
\author[a,2]{J.J.~G\'omez-Cadenas,\note{NEXT spokesperson}}
\author[a]{V.~\'Alvarez,}
\author[b]{C.D.R.~Azevedo,}
\author[c]{F.I.G.~Borges,}
\author[a]{S.~C\'arcel,}
\author[a]{J.V.~Carri\'on,}
\author[d]{S.~Cebri\'an,}
\author[a]{A.~Cervera,}
\author[c]{C.A.N.~Conde,}
\author[a]{J.~D\'iaz,}
\author[e]{M.~Diesburg,}
\author[f]{R.~Esteve,}
\author[g]{L.M.P.~Fernandes,}
\author[b]{A.L.~Ferreira,}
\author[g]{E.D.C.~Freitas,}
\author[h]{A.~Goldschmidt,}
\author[i]{D.~Gonz\'alez-D\'iaz,}
\author[j]{R.M.~Guti\'errez,}
\author[k]{J.~Hauptman,}
\author[g]{C.A.O.~Henriques,}
\author[l]{J.A.~Hernando~Morata,}
\author[f]{V.~Herrero}
\author[m]{L.~Labarga,}
\author[a]{A.~Laing,}
\author[e]{P.~Lebrun,}
\author[a]{I.~Liubarsky,}
\author[a]{N.~L\'opez-March,}
\author[a]{D.~Lorca,}
\author[j]{M.~Losada,}
\author[l]{G.~Mart\'inez-Lema,}
\author[a]{A.~Mart\'inez,}
\author[n]{F.~Monrabal,}
\author[g]{C.M.B.~Monteiro,}
\author[f]{F.J.~Mora,}
\author[b]{L.M.~Moutinho,}
\author[a]{P.~Novella,}
\author[n]{D.~Nygren,}
\author[a]{B.~Palmeiro,}
\author[e]{A.~Para,}
\author[a]{M.~Querol,}
\author[a]{J.~Renner,}
\author[p]{L.~Ripoll,}
\author[a]{J.~Rodr\'iguez,}
\author[c]{F.P.~Santos,}
\author[g]{J.M.F.~dos~Santos,}
\author[a]{L.~Serra,}
\author[n]{D.~Shuman,}
\author[a]{A.~Sim\'on,}
\author[q]{C.~Sofka,}
\author[a]{M.~Sorel,}
\author[q]{T.~Stiegler,}
\author[f]{J.F.~Toledo,}
\author[p]{J.~Torrent,}
\author[r]{Z.~Tsamalaidze,}
\author[b]{J.F.C.A.~Veloso,}
\author[q]{R.~Webb,}
\author[q,3]{J.T.~White,\note{Deceased}}
\author[a]{N.~Yahlali,}
\author[j]{H.~Yepes-Ram\'irez}

\emailAdd{justo.martin-albo@ific.uv.es}

\affiliation[a]{
Instituto de F\'isica Corpuscular (IFIC), CSIC \& Universitat de Val\`encia\\
Calle Catedr\'atico Jos\'e Beltr\'an, 2, 46980 Paterna, Valencia, Spain}
\affiliation[b]{
Institute of Nanostructures, Nanomodelling and Nanofabrication (i3N), Universidade de Aveiro\\
Campus de Santiago, 3810-193 Aveiro, Portugal}
\affiliation[c]{
LIP, Departamento de F\'isica, Universidade de Coimbra\\
Rua Larga, 3004-516 Coimbra, Portugal}
\affiliation[d]{
Laboratorio de F\'isica Nuclear y Astropart\'iculas, Universidad de Zaragoza\\ 
Calle Pedro Cerbuna, 12, 50009 Zaragoza, Spain}
\affiliation[e]{
Fermi National Accelerator Laboratory\\ 
Batavia, Illinois 60510, USA}
\affiliation[f]{
Instituto de Instrumentaci\'on para Imagen Molecular (I3M), Universitat Polit\`ecnica de Val\`encia\\ 
Camino de Vera, s/n, Edificio 8B, 46022 Valencia, Spain}
\affiliation[g]{
LIBPhys, Physics Department, University of Coimbra\\
Rua Larga, 3004-516 Coimbra, Portugal}
\affiliation[h]{
Lawrence Berkeley National Laboratory (LBNL)\\
1 Cyclotron Road, Berkeley, California 94720, USA}
\affiliation[i]{
European Organization for Nuclear Research (CERN)\\
CH-1211 Geneva 23, Switzerland}
\affiliation[j]
{Centro de Investigaci\'on en Ciencias B\'asicas y Aplicadas, Universidad Antonio Nari\~no\\ 
Sede Circunvalar, Carretera 3 Este No.\ 47 A-15, Bogot\'a, Colombia}
\affiliation[k]{
Department of Physics and Astronomy, Iowa State University\\
12 Physics Hall, Ames, Iowa 50011-3160, USA}
\affiliation[l]{
Instituto Gallego de F\'isica de Altas Energ\'ias, Univ.\ de Santiago de Compostela\\
Campus sur, R\'ua Xos\'e Mar\'ia Su\'arez N\'u\~nez, s/n, 15782 Santiago de Compostela, Spain}
\affiliation[m]{
Departamento de F\'isica Te\'orica, Universidad Aut\'onoma de Madrid\\
Campus de Cantoblanco, 28049 Madrid, Spain}
\affiliation[n]{
Department of Physics, University of Texas at Arlington\\
Arlington, Texas 76019, USA}
\affiliation[p]{
Escola Polit\`ecnica Superior, Universitat de Girona\\
Av.~Montilivi, s/n, 17071 Girona, Spain}
\affiliation[q]{
Department of Physics and Astronomy, Texas A\&M University\\
College Station, Texas 77843-4242, USA}
\affiliation[r]{
Joint Institute for Nuclear Research (JINR)\\
Joliot-Curie 6, 141980 Dubna, Russia}
%

\abstract{NEXT-100 is an electroluminescent high-pressure xenon gas time projection chamber that will search for the neutrinoless double beta (\bbonu) decay of \Xe. The detector possesses two features of great value for \bbonu\ searches: energy resolution better than 1\% FWHM at the $Q$ value of \Xe\ and track reconstruction for the discrimination of signal and background events. This combination results in excellent sensitivity, as discussed in this paper. Material-screening measurements and a detailed Monte Carlo detector simulation predict a background rate for NEXT-100 of at most $4\times10^{-4}$~counts~keV$^{-1}$~kg$^{-1}$~yr$^{-1}$. Accordingly, the detector will reach a sensitivity to the \bbonu-decay half-life of $2.8\times10^{25}$~years (90\% CL) for an exposure of 100 $\mathrm{kg}\cdot\mathrm{year}$, or $6.0\times10^{25}$~years after a run of 3 effective years.}

\keywords{Double beta decay}
\arxivnumber{1511.09246}

\maketitle
\flushbottom


\section{Introduction} \label{sec:Introduction}
Neutrinoless double beta (\bbonu) decay is a hypothetical second-order weak process in which a nucleus of atomic number $Z$ and mass number $A$ transforms into its isobar with atomic number $Z+2$ emitting two electrons: 
\begin{equation}
\ce{^{A}_{Z}X} \to\ \ce{^{A}_{Z+2}X} + e^{-} + e^{-}\,.
\end{equation}
The discovery of this process would prove that neutrinos are Majorana particles ---\thinspace that is, identical to their antiparticles\thinspace--- and that total lepton number is not conserved in nature, two findings with far-reaching implications in particle physics and cosmology. First of all, Majorana neutrinos imply the existence of a new energy scale at a level inversely proportional to the observed neutrino masses \cite{Weinberg:1979sa}. Such a scale, besides providing a simple explanation for the striking lightness of neutrino masses \cite{Minkowski:1977sc, GellMann:1980vs, Yanagida:1979as, Mohapatra:1979ia}, is probably connected to several open questions in particle physics, like the origin of mass or the flavour problem (see, for instance, Ref.~\cite{King:2015jkh}). Second, Majorana neutrinos violate the conservation of lepton number, and this, together with CP violation, could be responsible for the observed cosmological asymmetry between matter and antimatter through the mechanism known as \emph{leptogenesis} \cite{Fukugita:1986hr}.

Experimentally, no compelling evidence of the existence of \bbonu\ decay has been obtained so far. However, a new generation of experiments that are already running or about to run promises to push forward the current limits exploring the degenerate-hierarchy region of neutrino masses \cite{GomezCadenas:2011it, Giuliani:2012zu, Elliott:2012sp, Cremonesi:2013vla}. In order to do that, these experiments are using source masses ranging from tens to hundreds of kilograms and improving the background rates achieved by previous experiments by, at least, an order of magnitude. If no signal is found, masses in the tonne scale and further background reduction will be required to continue the exploration. Only a few of the techniques considered at present can possibly be extrapolated to those levels.

The \emph{Neutrino Experiment with a Xenon TPC} (NEXT)\footnote{\url{http://next.ific.uv.es/}} seeks to discover the neutrinoless double beta decay of \Xe\ using an electroluminescent time projection chamber filled with 100~kg of isotopically enriched xenon gas. This detector, named NEXT-100, possesses two features of great value for \bbonu-decay searches: excellent energy resolution ($<$1\% FWHM at 2.5~MeV) \cite{Alvarez:2012kua, Lorca:2014sra} and charged-particle tracking for the active suppression of background \cite{Alvarez:2013gxa, Ferrario:2015ina}. Furthermore, the technology can be extrapolated to large source masses, thus allowing the full exploration of the inverted-hierarchy region of neutrino masses. The installation and commissioning of NEXT-100 at the \emph{Laboratorio Subterr\'aneo de Canfranc} (LSC), Spain, is planned for 2018. Prior to that, the NEXT Collaboration will operate underground the NEW detector, a technology demonstrator that implements in a smaller scale the design chosen for NEXT-100 using the same materials and photosensors.

In this paper, we study the sensitivity of NEXT-100 to neutrinoless double beta decay. Some of the factors on which the detector's sensitivity depends are fixed by design, such as the isotope chosen (\Xe) or the available source mass (100 kg of xenon enriched to 91\% in \Xe). Other factors ---\thinspace the energy resolution or the tracking performance, for instance\thinspace---, can be extrapolated from results of the R\&D phase of the project. Finally, the levels of the potential backgrounds and the discrimination power of the detector can be estimated with information on the radiopurity of the construction materials and the use of Monte Carlo simulation. These last two factors ---\thinspace the background model and the discrimination capabilities of NEXT\thinspace--- are the focus of this paper.

\section{Rate of neutrinoless double beta decay} \label{sec:RateNeutrinoless}
Any source of lepton number violation can, in principle, induce neutrinoless double beta decay and contribute to its rate \cite{Feinberg:1968, Pontecorvo:1968wp, Mohapatra:1980yp, Mohapatra:1986su, Hirsch:1995vr, Tello:2010am}. In the simplest case, however, \bbonu\ decay is mediated by the virtual exchange of a light Majorana neutrino \cite{Racah:1937qq, Furry:1939qr}, and its rate is then given by
\begin{equation}
\left( T^{0\nu}_{1/2} \right)^{-1} = G^{0\nu} \left| M^{0\nu} \right|^{2} \left(\frac{\mbb}{m_{e}} \right)^{2}. \label{eq:RateNeutrinoless}
\end{equation}
Here, $G^{0\nu}$ is a \emph{phase-space factor} that depends on the energy release of the decay and on the nuclear charge $Z$, $M^{0\nu}$ is the \emph{nuclear matrix element} (NME) of the process, that is, a measure of the nuclear-structure aspects affecting the decay, $m_{e}$ is the electron mass, and \mbb\ is the so-called \emph{effective Majorana mass} of the electron neutrino:
\begin{equation}
\mbb \equiv \left| \sum_{i}\ U^{2}_{ei}\ m_{i} \right|\,, \label{eq:mbb}
\end{equation}
where $U_{ei}$ are the elements of the first row of the neutrino mixing matrix and $m_{i}$ are the neutrino mass eigenstates.

\begin{table}
\centering
\begin{tabular}{ll}
\toprule
Relative atomic mass & $135.907219(8)$ \cite{Wang:2012ame} \\ \midrule
$Q$ value $^{136}\mathrm{Xe}\to{^{136}\mathrm{Ba}}$ & 2457.83(37) keV \cite{Redshaw:2007un}  \\
          & 2458.7(6) keV \cite{McCowan:2010zz} \\
          & 2458.1(3) keV (average) \\ \midrule
$G^{0\nu}$ ($10^{-15}$~year$^{-1}$) & 14.58 \cite{Kotila:2012zza} \\
                                    & 14.54 \cite{Mirea:2014dza} \\ \midrule
\bbonu\ decay NME & 2.19 (ISM)   \cite{Menendez:2008jp} \\
                  & 3.05 (IBM-2) \cite{Barea:2015kwa} \\
                  & 2.46 (QRPA)  \cite{Simkovic:2013qiy} \\
                  & 2.91 (QRPA)  \cite{Hyvarinen:2015bda} \\
                  & 4.12 (EDF)   \cite{Vaquero:2014dna} \\
                  & 4.32 (EDF)   \cite{Yao:2014uta} \\
\bottomrule
\end{tabular}
\caption{Properties of \Xe\ relevant to neutrinoless double beta decay searches: relative atomic mass, $Q$ value of the decay (i.e.\ mass difference between the parent and daughter atoms), phase-space factor ($G^{0\nu}$) and nuclear matrix element (NME). The figures in parentheses after the first two quantities give the 1$\sigma$ experimental uncertainty in the last digits. The uncertainties on the $G^{0\nu}$ calculations (originating from the uncertainties on the $Q$ value and the nuclear radius) are of the order of 5--10\% \cite{Kotila:2012zza}. The quoted nuclear matrix elements (NME) are the most recent calculations for \bbonu\ decay to the ground state mediated by light-neutrino exchange in four different nuclear-theory frameworks: \emph{interacting shell model} (ISM), \emph{interacting boson model} (IBM-2), \emph{quasiparticle random-phase approximation} (QRPA) and \emph{energy density functional theory} (EDF). All NMEs are dimensionless and have been calculated with the free-nucleon value of the axial-vector coupling constant ($g_{A}\simeq1.26$), with model uncertainties varying between 15 and 30\%.} \label{tab:Xe136}
\end{table}

While the phase-space factor can be computed analytically with high accuracy \cite{Kotila:2012zza, Mirea:2014dza}, only approximate estimates of the NME can be obtained at present due to the many-body nature of the nuclear problem. Table~\ref{tab:Xe136} lists the most recent calculations of the NME of \Xe\ for \bbonu\ decay mediated by light-neutrino exchange from a variety of nuclear models. The results are not completely convergent, differing by up to a factor of 2. An even more significant source of uncertainty results from the dependence of the NME on the square of the axial-vector coupling constant, $g_{A}$. While the calculated NMEs for \bbonu\ decay are generally presented with the free-nucleon value ($g_{A}\simeq1.26$), in the case of the standard double beta decay with neutrinos (\bbtnu), consisting of two simultaneous beta decays,
\begin{equation}
\ce{^{A}_{Z}X} \to\ \ce{^{A}_{Z+2}X} + e^{-} + e^{-} + \overline{\nu}_{e} + \overline{\nu}_{e}\,,
\end{equation}
effective (or quenched) values of $g_{A}$ of about 0.6--0.8 (depending on the NME calculation framework) \cite{Barea:2015kwa} are required to match the measured half-lives \cite{Barabash:2015eza}. The difference between 0.6 and 1.26 translates into a factor of 20 in rate. The extent of the quenching in \bbonu\ decay ---\thinspace whether or not it is the same as in \bbtnu\ decay\thinspace--- is under debate among nuclear theorists \cite{Barea:2013bz, Engel:2014pha, Engel:2015wha, Dell'Oro:2016dbc}.

\begin{figure}
\centering
\includegraphics[width=0.5\textwidth]{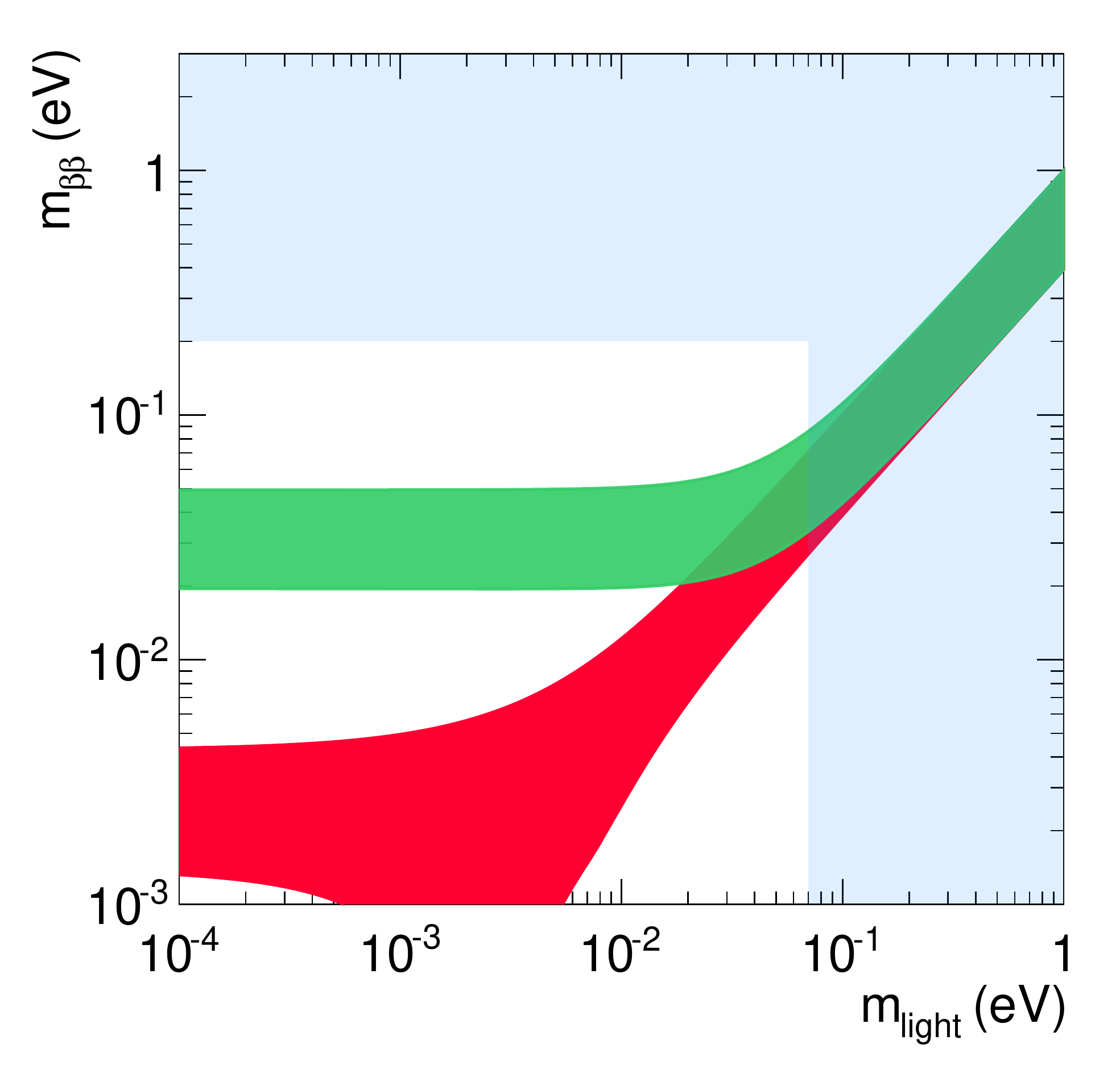}
\caption{The effective Majorana neutrino mass, \mbb, as a function of the lightest neutrino mass, $m_\mathrm{light}$. The green band corresponds to the inverted ordering of neutrino masses, while the red band corresponds to the normal ordering. The horizontally-excluded region comes from experimental bounds on the half-life of \bbonu\ decay \cite{Agostini:2013mzu, Albert:2014awa, Asakura:2014lma}; the vertical one, from cosmological constraints on $m_\mathrm{light}$ \cite{Ade:2015xua}.} \label{fig:NuMajoranaMassVsNuLight}
\end{figure}

Equations (\ref{eq:RateNeutrinoless}) and (\ref{eq:mbb}) show that a measurement of the \bbonu\ decay rate would provide direct information on neutrino masses (under the assumption of light neutrino exchange), albeit with important uncertainties from nuclear and neutrino physics. The relationship between \mbb\ and the lightest neutrino mass is shown in Figure~\ref{fig:NuMajoranaMassVsNuLight}. The width of the allowed bands is due to the unknown CP violation phases and the 3$\sigma$ uncertainties in the mixing parameters measured in neutrino oscillation experiments \cite{Gonzalez-Garcia:2014bfa}. The figure also shows an upper bound on the lightest neutrino mass from cosmological observations ($m_\mathrm{light}<0.23/3$~eV) \cite{Ade:2015xua} and the current upper bound on \mbb\ from \bbonu-decay searches ($\mbb<0.2$~eV) \cite{Agostini:2013mzu, Albert:2014awa, Asakura:2014lma}.

\section{Experimental sensitivity to $m_{\beta\beta}$} \label{sec:Sensitivity}

The detectors used for double beta decay searches are designed to measure the energy of the radiation emitted by a \bb\ source. In a \bbonu\ decay, the sum of the kinetic energies of the two released electrons is always equal to the $Q$ value of the process, that is, the mass difference between the parent and daughter atoms: 
\begin{equation}
Q_{\beta\beta} \equiv M(\ce{^{A}_{Z}X}) - M(\ce{^{A}_{Z+2}X})\,.
\end{equation}
In practice, due to the finite energy resolution of any detector, \bbonu\ events would spread over an energy range centred around \Qbb\ following, typically, a Gaussian distribution. Other processes occurring in the detector can fall within that energy window becoming a background. If a \bbonu\ peak were observed in an experiment, the number of signal events could then be related to the half-life of the process as follows:
\begin{equation}
T^{0\nu}_{1/2} = \log2~\frac{N_\mathrm{A}}{W}~\frac{\varepsilon~M~t}{N}\,, \label{eq:HalfLife}
\end{equation}
where $N_\mathrm{A}$ is the Avogadro constant, $W$ is the atomic mass of the \bbonu-decaying isotope, $\varepsilon$ is the signal detection efficiency, $M$ is the source mass, $t$ is the measuring time and $N$ is the number of \bbonu\ events observed in the experiment. 

Accordingly, combining Eqs.~(\ref{eq:RateNeutrinoless}) and (\ref{eq:HalfLife}), we find that the sensitivity to \mbb\ of an experiment searching for \bbonu\ decay is
\begin{equation}
\mathcal{S}(\mbb) = \mathcal{K}~\sqrt{\frac{\overline{N}}{\varepsilon\,M\,t}}\ , \label{eq:Sensmbb}
\end{equation}
where 
\begin{equation}
\mathcal{K} \equiv \left(\frac{W~m_{e}^{2}}{\log2~N_\mathrm{A}~G^{0\nu}~\left|M^{0\nu}\right|^{2}}\right)^{1/2}
\end{equation}
is a constant that depends solely on the source isotope employed and $\overline{N}$ is the average upper limit on the number of events expected in the experiment under the no-signal hypothesis. For an experiment with Poisson-distributed background of mean $b$, the average upper limit is given by
\begin{equation}
\overline{N}(b) = \sum^{\infty}_{n=0}\ \frac{b^{n}\,e^{-b}}{n!}~\mathcal{U}(n|b)\,,
\end{equation}
where $\mathcal{U}(n|b)$ is a function that returns a frequentist upper limit ---\thinspace Feldman-Cousins \cite{Feldman:1997qc}, for instance\thinspace--- at a certain confidence level for a given observation $n$ and a known expected value $b$.

For the case of high background, the average upper limit is proportional to the square root of the mean number of background events \cite{GomezCadenas:2010gs}: $\overline{N}\propto\sqrt{b}\,$. Besides, the number of background events is usually proportional to the exposure, $Mt$, and to the width of the energy window defined by the resolution of the detector, $\Delta E$: $b=c\cdot M\cdot t\cdot \Delta E$, where $c$ is the expected background rate, typically expressed in counts~keV$^{-1}$~kg$^{-1}$~yr$^{-1}$. Substituting these two expressions into Eq.~(\ref{eq:Sensmbb}), we obtain a well-known figure of merit for \bbonu-decay experiments:
\begin{equation}
\mathcal{S}(\mbb) \propto \sqrt{1/\varepsilon}~\left( \frac{c~\Delta E}{M\,t} \right)^{1/4}\,. \label{eq:FigureOfMerit}
\end{equation}
The above formula shows that the presence of background in the region of interest around \Qbb\ limits considerably the sensitivity of an experiment, improving only as $(M\,t)^{-1/4}$ instead of the inverse square-root dependence, Eq.~(\ref{eq:Sensmbb}), expected in the case of negligible background.

\section{The NEXT experiment} \label{sec:Next100}
NEXT will search for the \bbonu\ decay of \Xe\ making use of a 100-kg xenon gas TPC with electroluminescent amplification and optical readouts. Xenon is a good detection medium that provides strong scintillation and ionization signals. Moreover, in its gaseous phase, xenon offers very good energy resolution; better in principle than 0.5\% FWHM at the $Q$ value of \Xe\ \cite{Nygren:2009zz}. In order to achieve optimal resolution, the ionization signal is amplified in NEXT using the \emph{electroluminescence} (EL) of xenon: the electrons liberated by ionizing particles passing through the gas drift towards the TPC anode under the influence of a moderate electric field (0.3--0.5~kV~cm$^{-1}$), entering then into another region where they are accelerated by a stronger field (2--3~kV~cm$^{-1}$~bar$^{-1}$), intense enough so that the electrons can excite the Xe atoms but not enough to ionize them. This excitation energy is ultimately released, with sub-Poissonian fluctuations, in the form of proportional secondary scintillation light (or EL). An array of photomultiplier tubes (PMTs) ---\thinspace the so-called \emph{energy plane}\thinspace--- located behind the TPC cathode detects a fraction of these EL photons to provide a precise measurement of the total energy deposited in the gas. These PMTs detect as well the primary scintillation, which is used to mark the start of the event ($t_{0}$). The forward-going EL photons are detected by a dense array of silicon photomultipliers (SiPMs) ---\thinspace the \emph{tracking plane}\thinspace--- located behind the anode, very close to the EL region, and the associated signals are used for track reconstruction.

The initial phase of the NEXT experiment was devoted to the demonstration of the detector concept described above using two prototypes, NEXT-DEMO and NEXT-DBDM, that contained approximately 1~kg of natural xenon at 10--15~bar. An energy resolution of 1\% FWHM at 662~keV, which ---\thinspace assuming a $1/\sqrt{E}$ dependence\thinspace--- extrapolates to 0.5\% FWHM at the $Q$ value of \Xe, was measured in the DBDM prototype \cite{Alvarez:2012kua}. The best resolution measured in DEMO, 1.62\% FWHM at 511~keV, extrapolates to 0.74\% FWHM \cite{Lorca:2014sra}. In addition, the NEXT-DEMO prototype has shown that track reconstruction is possible with an EL-based amplification scheme \cite{Alvarez:2013gxa} and that the reconstructed energy-deposition pattern can be used for the identification of signal-like and background-like event topologies \cite{Ferrario:2015ina}. 

The current stage of the NEXT project involves the operation at LSC of the {\scshape NEXT-White}\footnote{Named after our late collaborator Prof.\ James T.\ White.} (NEW) detector, a technology demonstrator that implements in a 1:2 scale the design chosen for the 100-kg detector (NEXT-100) using the same materials and photosensors. The NEW data will make possible the optimization of calibration and reconstruction methods and the validation of the NEXT-100 background model. The NEXT-100 detector, described in some detail in the remainder of this section, is planned to start taking low-background data in 2018.

\subsection{NEXT-100}
\begin{figure}
\centering
\includegraphics[trim=50 50 50 50, clip, width=0.9\textwidth]{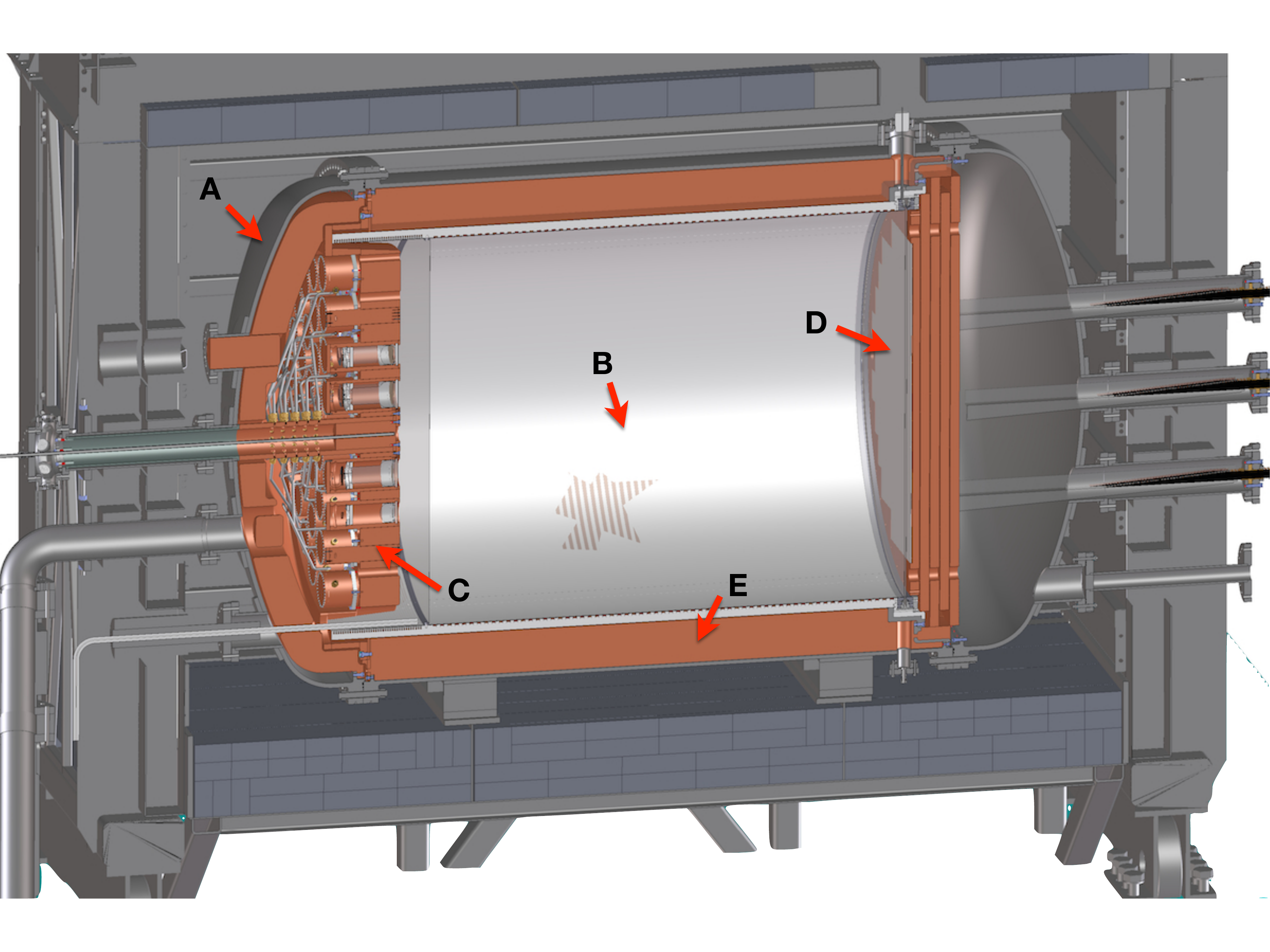}
\caption{Cross-section view of the NEXT-100 detector inside its lead castle shield. A stainless-steel pressure vessel (A) houses the electric-field cage (B) and the two sensor planes (energy plane, C; tracking plane, D) located at opposite ends of the chamber. The active volume is shielded from external radiation by at least 12~cm of copper (E) in all directions.} \label{fig:Next100}
\end{figure}

Figure~\ref{fig:Next100} shows a longitudinal cross-section schematic of NEXT-100. The active volume of the detector is a cylinder of approximately 1.15~m$^{3}$ that can hold about 100 kg of xenon gas at 15~bar. It is surrounded by a series of copper rings for electric-field shaping that are fixed to the inner surface of an open-ended high-density polyethylene (HDPE) cylindric shell, 2.5~cm thick, 148~cm long and 107.5~cm in diameter, that  provides structural stiffness and electric insulation. The rings are covered by polytetrafluoroethylene (PTFE) tiles coated with tetraphenyl-butadiene (TPB) to shift the xenon VUV light to the blue region (around 440~nm) so as to improve the light collection efficiency. One of the ends of the HDPE cylinder is closed by a fused-silica window 1 cm thick. This window functions as the TPC anode thanks to a transparent, conductive, wavelength-shifting coating of indium tin oxide (ITO) and TPB. The two other electrodes of the TPC, EL gate and cathode, are positioned 0.5~cm and 106.5~cm away from the anode, respectively. They are built with highly transparent stainless steel wire mesh stretched over circular frames. The electrodes will be set at voltages such that a moderate electric field of 0.3--0.5~kV~cm$^{-1}$ is established in the drift region between cathode and gate, and another field of higher intensity, 2--3~kV~cm$^{-1}$~bar$^{-1}$, is created in the EL gap, between gate and anode, for the amplification of the ionization signal. The high voltage is supplied to the electrodes via radiopure, custom-made feed-throughs. 

The energy plane of NEXT-100 will be composed of 60 Hamamatsu R11410-10 photomultiplier tubes located behind the cathode of the TPC and covering approximately 30\% of its area. This coverage is a compromise between the need to collect as much light as possible for a robust measurement of the energy and $t_{0}$, and the need to minimize the number of sensors to reduce cost, technical complexity and radioactivity. The R11410-10 is a 3-in PMT specially developed for low-background operation \cite{Lung:2012pi}. It is equipped with a synthetic silica window and a photocathode made of low temperature bialkali with quantum efficiency above 30\% for the emission wavelengths of xenon and TPB \cite{Lung:2012pi}. Pressure-resistance tests run by the manufacturer showed that the R11410-10 cannot withstand pressures above 6 atmospheres \cite{Lung:2012pi}. Therefore, in NEXT-100 they will be sealed into individual pressure-resistant, vacuum-tight copper enclosures closed with sapphire windows 5~mm thick. The PMTs are optically coupled to the windows using an optical gel with a refractive index intermediate between those of fused silica and sapphire. The external face of the enclosure windows is coated with TPB. The enclosures are all connected via vacuum-tight tubing conduits to a central manifold and maintained at vacuum. The PMT cables route through the conduits and the central manifold to a feedthrough in the pressure vessel nozzle. 

The tracking function in NEXT-100 will be provided by an array of 7168 SiPMs regularly positioned at a pitch of 1~cm and located behind the fused-silica window that closes the EL gap. The SiPMs, manufactured by {\scshape SensL}, have an active area of 1 mm$^{2}$, sensitive cells of 50 $\mu$m size and high photon detection efficiency in the blue region (about 40\% at 440~nm). They are very cost-effective and their radioactivity is very low, given their composition and small mass. The SiPMs will be mounted on flexible circuit boards made of Kapton and copper, each one with 64 sensors arranged as an $8\times8$ matrix. The boards have long tails that carry the signals through zigzagging slits ---\thinspace so as to avoid a straight path for external gammas\thinspace--- made in the copper plates that shield the active volume. The tails are connected to flat shielded cables that extract the signals from the vessel via large custom-made feed-throughs. 

The sensor planes and the electric-field cage are contained within a stainless-steel pressure vessel that consists of a cylindrical central shell of 160~cm length, 136~cm inner diameter and 1~cm wall thickness, and two identical torispherical heads of 35~cm height, 136~cm inner diameter and 1~cm wall thickness. It has been fabricated with stainless steel Type 316Ti (acquired from Nironit) due to its low levels of natural radioactive contaminants. Designed almost entirely by the Collaboration following the ASME Pressure Vessel Code, the vessel has been built by a specialized company based in Madrid. The field cage is surrounded by a set of 12-cm thick copper bars parallel to the TPC symmetry axis, and both sensor planes are mounted to copper plates of 12~cm thickness attached to internal flanges of the vessel heads. The active volume of the detector is, therefore, shielded from external radiation by at least 12~cm of copper in all directions. The vessel sits on top of an anti-seismic pedestal and inside of a 20-cm thick lead shield made of staggered lead bricks held by a stainless-steel frame.

\section{Sources of background in NEXT} \label{sec:SignalAndBackground}
\begin{figure}
\centering
\includegraphics[width=0.95\textwidth]{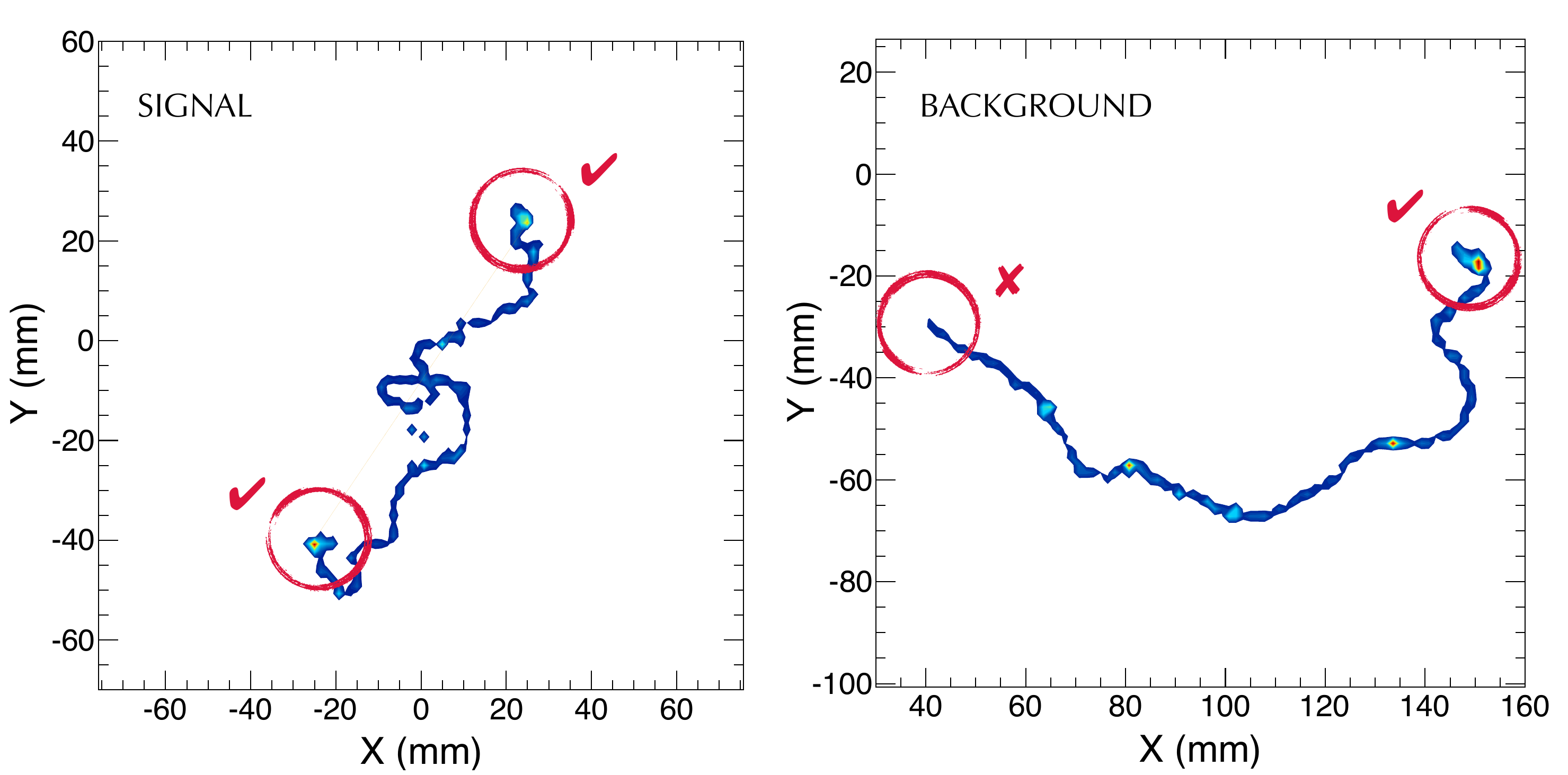}
\caption{Monte Carlo simulation of signal (\bbonu\ decay of \Xe) and background (single electron of energy equal to the $Q$ value of \Xe) events in gaseous xenon at 15~bar. The colour scale codes the energy loss per path length. The ionization tracks left by signal events feature large energy deposits (or \emph{blobs}) at both ends.} \label{fig:TrackSignature}
\end{figure}

In this section we discuss the various potential components of the background spectrum of NEXT-100 in the energy region around \Qbb. The relevance of any background source in NEXT depends on its probability to generate a signal-like track in the active volume of the detector with energy around the $Q$ value of \Xe. In principle, charged particles (muons, betas, etc.) entering the detector can be eliminated with high efficiency ($>99\%$) by defining a small veto region (of a few centimetres) around the boundaries of the active volume. Electric field inhomogeneities or malfunctioning photosensors could affect negatively this performance, but those effects can be measured with periodic calibrations using, for instance, crossing muons. Confined tracks generated by external neutral particles (such as high-energy gamma rays) or by internal contamination in the xenon gas can be suppressed taking advantage of the distinctive energy-deposition pattern of signal events, illustrated in Fig.~\ref{fig:TrackSignature}. Below the so-called \emph{critical energy} (about 12~MeV in gaseous xenon \cite{Agashe:2014kda}), electrons (and positrons) lose their energy at a relatively fixed rate until they become non-relativistic. At about that time, their effective $\mathrm{d}E/\mathrm{d}x$ rises, mostly due to the occurrence of strong multiple scattering, and the particles lose the remainder of their energy in a relatively short distance generating a \emph{blob}. Double beta decay events consist of two electrons emitted from a common vertex. Their reconstructed tracks, therefore, feature blobs at both ends. Background tracks, in contrast, are generated mostly by single electrons, thus having only one end-of-track blob.

\subsection{High-energy gamma rays from the natural decay series}
Natural radioactivity in detector materials and surroundings is, as in most other \bbonu-decay experiments, the main source of background in NEXT. In particular, the hypothetical \bbonu\ peak of \Xe\ ($\Qbb=2458.1\pm0.3$~keV \cite{Redshaw:2007un, McCowan:2010zz}) lies in between the photo-peaks of the high-energy gammas emitted after the $\beta$ decays of \Bi\ and \Tl, intermediate products of the uranium and thorium series, respectively: 
\begin{itemize}
\item The daughter isotope of \Bi, $^{214}$Po, emits a number of de-excitation gammas with energies around and above the $Q$ value of \Xe\ \cite{Wu:2012nds}. The gamma line at 2447~keV (1.57\% intensity \cite{Wu:2012nds}) is very close to \Qbb, and its photoelectric peak would overlap the signal peak even for energy resolutions as good as 0.5\% FWHM. All the other gamma lines emitted after the decay of \Bi\ have very low intensity (at least two orders of magnitude lower than the 2447~keV line), and hence their contribution to the background rate can be neglected. 
\item The decay product of \Tl, $^{208}$Pb, emits a de-excitation photon of 2615~keV with an intensity of 99.75\% \cite{Martin:2007nds}. Electron tracks from its photo-peak can lose energy via bremsstrahlung and fall in the \emph{region of interest} (ROI) around \Qbb\ defined by the energy resolution of the detector. Additionally, even though the Compton edge of the 2.6~MeV gamma is at 2382~keV, well below \Qbb, the Compton-scattered photon can generate other electron tracks close enough to the initial Compton electron to be reconstructed as a single track with energy around \Qbb. 
\end{itemize}


\begin{table}
\centering
{\small
\begin{tabular}{l l l l D{.}{.}{2.9} D{.}{.}{2.9}}
\toprule
Material & Subsystem & Technique & Units & \multicolumn{1}{l}{\Tl} & \multicolumn{1}{l}{\Bi} \\ \midrule
Copper CuA1 & IS, EP, FC & GDMS & mBq/kg & <0.0014 & <0.012 \\
Fused silica & FC & NAA & mBq/kg & 0.034(4) & 0.21(5) \\
Kapton board & TP & HPGe & mBq/unit & 0.0104(11) & 0.070(5) \\
Lead & OS & GDMS & mBq/kg & 0.034(7) & 0.35(7) \\
PMT R11410-10 & EP & HPGe & mBq/PMT & 0.19(4) & 0.35(8) \\
Polyethylene & FC & ICPMS & mBq/kg & <0.0076 & <0.062 \\
Resistor (1~G$\Omega$) & FC & HPGe & mBq/unit & 0.000011(6) & 0.00009(4) \\
Sapphire & EP & NAA & mBq/kg & 0.04(1) & <0.31 \\
Steel 316Ti & PV & GDMS, HPGe & mBq/kg & <0.15 & <0.46 \\
SiPM SensL & TP & HPGe & mBq/unit & <0.00003 & <0.00009 \\
\bottomrule
\end{tabular} }
\caption{Specific activity of \Tl\ and \Bi\ in the most relevant materials and components used in the NEXT-100 detector \cite{Alvarez:2012as, Dafni:2014yja, Alvarez:2014kvs, Cebrian:2015jna}. The figures in parentheses after the measurements give the 1-standard-deviation uncertainties in the last digits; the limits are given at 95\% CL. Three items (fused silica, sapphire and the field-cage resistors) have not been screened yet with sufficient precision; therefore, we use instead  measurements by the EXO Collaboration \cite{Leonard:2007uv, Auger:2012gs}. The activities determined via mass spectrometry (GDMS or ICPMS) or neutron activation analysis (NAA) were derived from Th and U concentrations. High-purity germanium (HPGe) $\gamma$-ray spectroscopy results correspond, whenever possible, to the lower parts of the natural decay chains.  The abbreviations used to refer to the NEXT-100 detector subsystems have the following meaning: EP stands for energy plane; TP, for tracking plane; FC, for electric-field cage; PV, for pressure vessel; IS, for inner shielding; and OS, for outer shielding.} \label{tab:ActivityMaterials}
\end{table}



\begin{table}[!]
\centering
\begin{tabular}{lll c c}
\toprule
Detector subsystem & Material & Quantity & \multicolumn{1}{c}{\Tl} & \Bi\ \\ 
                   &          &          & \multicolumn{1}{c}{(mBq)} & (mBq) \\ \midrule
\emph{Pressure vessel} \\
\quad Total & Steel 316Ti & 1310~kg & $<197$ & $<603$ \\ \addlinespace
\emph{Energy plane} \\
\quad PMTs & R11410-10 & 60~units & $11(3)$ & $21(5)$ \\
\quad PMT enclosures & Copper CuA1 & 60$\times$4.3 kg & $<0.36$ & $<3.1$ \\
\quad Enclosure windows & Sapphire & 60$\times$0.14 kg & $0.34(8)$ & $<2.6$ \\ \addlinespace
%
\emph{Tracking plane} \\
\quad SiPMs & {\scshape Sensl} 1~mm$^{2}$ & 107$\times$64 units & $<0.2$ & $<0.6$ \\
\quad Boards & Kapton FPC & 107 units & $1.11(12)$ & $7.5(5)$ \\ \addlinespace 
\emph{Field cage} \\
\quad Barrel & Polyethylene & 128~kg & $<1$ & $<8$ \\
\quad Shaping rings & Copper CuA1 & 120$\times$3~kg & $<0.5$ & $<4$ \\
\quad Electrode rings & Steel 316Ti & 2$\times$5~kg & $<1.5$ & $<5$ \\
\quad Anode plate & Fused silica & 9.5~kg & $0.32(4)$ & $2.0(5)$ \\
\quad Resistor chain & 1-G$\Omega$ resistors & 240 units & $<0.0026$ & $<0.02$ \\ \addlinespace
\emph{Shielding} \\
\quad Inner shield & Copper CuA1 & 9620~kg & $<13$ & $<120$ \\
\quad Outer shield & Lead & 60700~kg & $2060(430)$ & 21300(4300) \\
\bottomrule
\end{tabular}
\caption{Radioactivity budget of the NEXT-100 detector. The figures in parentheses after the measurements give the 1-sigma uncertainties in the last digits. The upper limits in the activity of most subsystems originate in the 95\% CL limits set on the specific activity of the corresponding materials quoted on Table~\ref{tab:ActivityMaterials}.} \label{tab:RadioactivityBudget}
\end{table}

The NEXT Collaboration is carrying out a thorough campaign of material screening and selection using gamma-ray spectroscopy (with the assistance of the LSC Radiopurity Service) and mass spectrometry techniques (ICPMS and GDMS). Table~\ref{tab:ActivityMaterials} collects the measurements of the specific activity of \Tl\ and \Bi\ in the most relevant materials and components used in the NEXT-100 detector, and Table~\ref{tab:RadioactivityBudget} details the radioactivity budget of NEXT-100 separated into detector subsystems.

The rock walls of the underground laboratory are a rather intense source of high-energy gammas due to the presence of trace radioactive contaminants in their composition. The total gamma flux at LSC (Hall A) is $1.06\pm0.24$~cm$^{-2}$~s$^{-1}$, with contributions from $^{40}$K ($0.52\pm0.23$~cm$^{-2}$~s$^{-1}$), $^{238}$U ($0.35\pm0.03$~cm$^{-2}$~s$^{-1}$) and $^{232}$Th ($0.19\pm0.04$~cm$^{-2}$~s$^{-1}$) \cite{Bettini:2012fu,Bandac:2014}. Nevertheless, the external lead shield of NEXT-100 will attenuate this flux by more than 4 orders of magnitude, making its contribution to the final background rate negligible compared to that of the natural radioactivity from detector construction materials. 

\subsection{Radon}
Radon, another intermediate decay product of the uranium and thorium series, is a concern for most \bbonu-decay experiments. It is one of the densest substances that remains a gas under normal conditions, and it is also the only gas in the atmosphere that solely has radioactive isotopes. Being a noble gas, radon is chemically not very reactive and can diffuse easily through many materials infiltrating into the active region of the detectors. While the average rate of production of $^{220}$Rn (from the thorium decay series) is about the same as $^{222}$Rn (from the uranium decay series), the longer half-life of the latter (3.8 days versus 55 seconds) makes it, normally, much more abundant. Radon progeny, also radioactive, tend to be charged and adhere to surfaces or dust particles.

The measured activity of airborne radon ($^{222}$Rn) at the Laboratorio Subterr\'aneo de Canfranc (Hall A) varies between 60 and 80~Bq~m$^{-3}$ \cite{Bandac:2013lsc}. Left at this level, radon would represent an intolerably high source of gamma rays from \Bi. For this reason, the vicinity of the detector (the internal volume of the lead castle shield) will be flushed with clean air produced by a radon mitigation system such as those used, for instance, by the NEMO-3 \cite{Nachab:2007zz} and DarkSide \cite{Bossa:2014cfa} experiments. A reduction of, at least, a factor of 100 in the activity of airborne radon is expected.

Radon can also emanate from detector components and be transported to the active volume through the gas circulation. The $\alpha$ decays of radon (either $^{220}$Rn or $^{222}$Rn) in the bulk xenon do not represent a background: they have energies well above \Qbb\ and their very short tracks are easily identified \cite{Alvarez:2012hu, Serra:2014zda}. These alphas are useful, however, to monitor the activity of radon in the xenon gas \cite{Serra:2014zda, Albert:2013gpz}. The progeny of radon is positively charged and will drift toward the TPC cathode. A majority of the subsequent \Bi\ and \Tl\ beta decays will occur on the cathode rather than in the active volume \cite{Albert:2013gpz, Albert:2015nta}. These cathode events are equivalent to other background sources close to the active volume (\Tl\ and \Bi\ decays from the sensor planes, for instance): if the $\beta$ particle enters the active volume, the event can then be vetoed; otherwise, the de-excitation gamma rays that interact in the xenon can generate background tracks. In addition, a small fraction (0.2\%) of the \Bi\ $\beta$ decays occurring in the xenon bulk will produce an electron track with energy around \Qbb. Luckily, the disintegration of \Bi\ is followed shortly after by the $\alpha$ decay of $^{214}$Po ($T_{1/2} = 164~\mu$s \cite{Wu:2012nds}). The detection of this so-called Bi-Po coincidence can be used to identify and suppress with high efficiency these background events.

The design of NEXT-100 minimizes the use of materials and components known to emanate radon in high rates, such as plastics, cables or certain seals. Nevertheless, estimating a priori the emanation rate and radon activity in the xenon is difficult, since the available data are scant and have been acquired in very different conditions (in vacuum, typically) to those of NEXT-100. The Collaboration has designed a radon trap for the gas circulation system in case the activity of radon in the xenon becomes too high.

\subsection{Muons and neutrons}
Muons (and neutrinos) are the only surviving radiation from the atmosphere and outer space at the depths of the underground laboratories. At the LSC, the muon flux (integrated over all angles) was measured to be about $5\times10^{-3}$~m$^{-2}$~s$^{-1}$ \cite{Luzon:2006sh}, with an estimated mean energy of single muons of a few hundreds of GeV \cite{Lipari:1991ut}. Thanks to the tracking capabilities of NEXT-100, muons crossing the active volume of the detector will be easily identified and rejected. Other muons, however, may generate electromagnetic showers in their interactions with the shielding and detector materials. Some of the resulting high-energy gammas can reach the active volume of NEXT-100 and generate electron tracks. According to our estimations, the event rate induced by this potential source of background appears to be largely negligible compared to natural radioactivity (at least two orders of magnitude smaller), and it could be further suppressed with the installation of a muon veto surrounding the lead shield.

Neutrons, produced by cosmic-ray-muon interactions and radioactive decays, are another possible background source. They can activate the enriched xenon and produce $^{137}$Xe, a $\beta$ emitter with a $Q$ value of 4173~keV and a half-life of 3.8 minutes \cite{Browne:2007nds}. A small fraction ($\sim$0.5\%) of the $\beta$ tracks would fall within the region of interest of the energy spectrum. Our initial estimates indicate that the copper shielding surrounding the active volume would suppress the neutron flux enough so as to make the rate from this background source at least two orders of magnitude lower than that of the radioactive background from \Tl\ and \Bi.

\section{Signal detection efficiency and background rejection in NEXT-100}
We describe in this section a basic set of selection criteria for \bbonu\ events detected and reconstructed in NEXT-100. Applying this event selection to simulation datasets discussed below, we have obtained an estimate of the signal detection efficiency and background rejection power of the NEXT-100 detector. This event selection will also serve as a benchmark for more sophisticated algorithms for the discrimination of signal and background that are currently under development.

\subsection{Simulation and reconstruction} \label{sec:Simulation}
The study presented here made use of large datasets ---\thinspace of the order of $10^{10}$ simulated events\thinspace--- produced with NEXUS \cite{Martin-Albo:2015dza}, a Geant4-based \cite{Agostinelli:2002hh} detector simulation developed by the NEXT Collaboration. The events were generated with the fast-simulation mode of NEXUS, which provides as output for each event a collection of three-dimensional hits representing the ionization tracks left by charged particles in the active volume of NEXT-100. The {\scshape Decay0} event generator \cite{Ponkratenko:2000um} was used for the production of 10 million events of the \bbtnu\ decay of \Xe\ and 1 million of the \bbonu\ decay. These events were read afterwards by the detector simulation program and given a random initial position within the xenon gas volume. Natural-radioactivity backgrounds were simulated using the radioactive-decay module included in Geant4. For each isotope and each detector subsystem considered a source of background in the detector (see Table~\ref{tab:RadioactivityBudget}), one billion events were simulated with initial positions uniformly distributed within the corresponding volume. In all cases, events with total energy deposited in the active volume below 2.3~MeV were discarded already at the simulation stage. Next, the active volume was divided into a regular grid of small cuboids ($1\times1\times1$~cm$^{3}$) known as \emph{voxels}, and the hits produced by the simulation were accumulated in the voxels that contained their positions. The total energy deposited in the active volume was smeared according to the detector's energy-resolution function (0.75\% FWHM at the Q value of \Xe); the energy accumulated in each voxel was multiplied by the ratio of the smeared energy to the original one. Both the size of the voxels and the energy resolution are extrapolations from results of the NEXT-DEMO prototype \cite{Lorca:2014sra, Ferrario:2015ina}.

\begin{figure}
\centering
\includegraphics[width=0.45\textwidth]{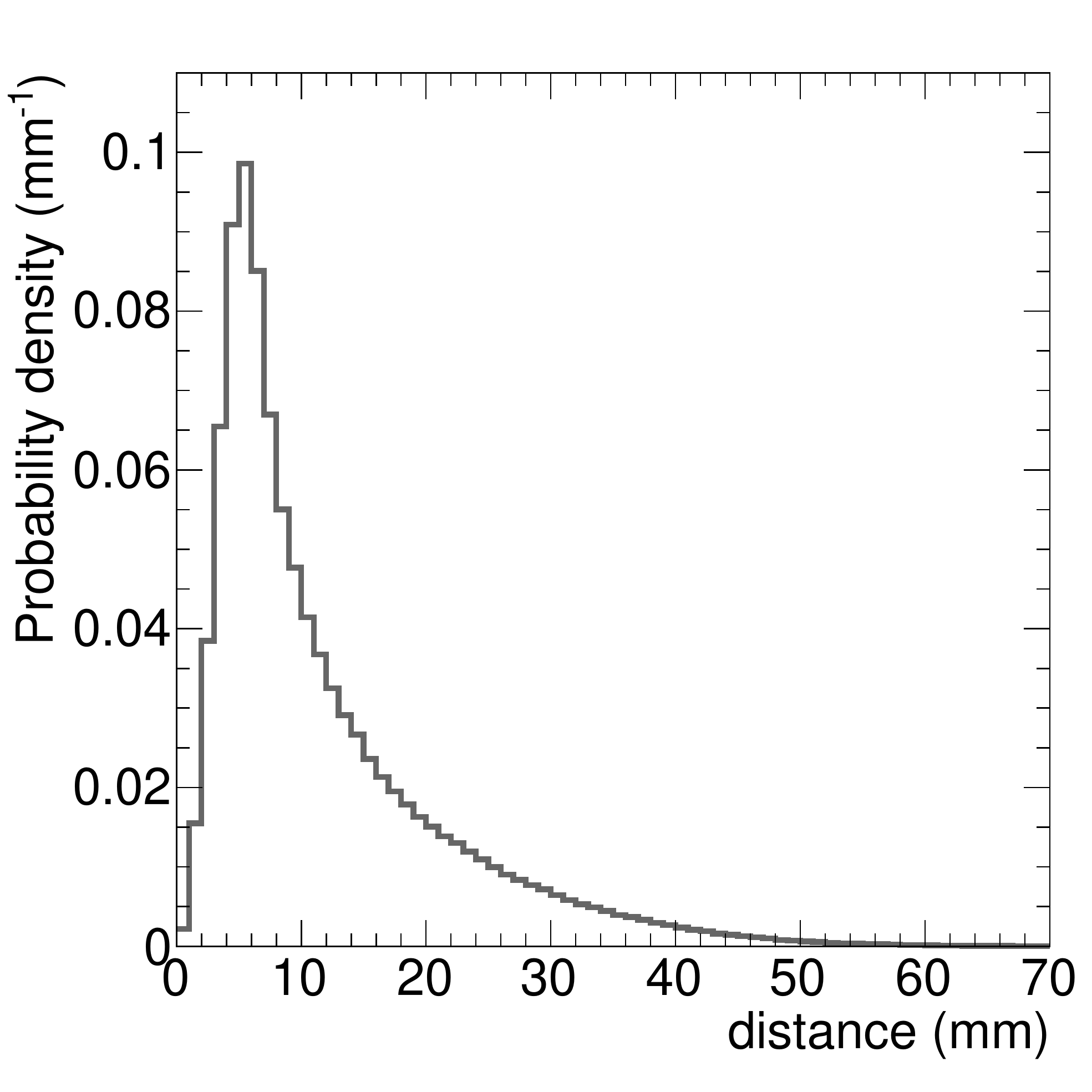}
\caption{Distance between the reconstructed (using a graph-search algorithm) ends of a track and the true ends for electron tracks of 2458~keV simulated with NEXUS.} \label{fig:Extremes}
\end{figure}

The collection of voxels resulting from each simulated event can be regarded as a \emph{graph}, that is, a set of \emph{nodes} and the \emph{links} that connect them. A graph of $n$ nodes can be characterized by its \emph{distance matrix}, a square ($n \times n$) matrix that contains the distance between any pair of nodes. Specifically, the matrix element $d_{ij}$, which indicates the distance between nodes $i$ and $j$, is defined in the following way:
\begin{itemize}
\item $d_{ij}=\sqrt{(x_{i}-x_{j})^{2} + (y_{i}-y_{j})^{2} + (z_{i}-z_{j})^{2}}$ if the cubic voxels corresponding to nodes $i$ and $j$ are in contact (i.e.\ they share one face, one edge or one vertex). For cubic voxels of 1~cm$^{3}$, like in our case, there are only three possible values: $1$, $\sqrt{2}$ and $\sqrt{3}$~cm.
\item $d_{ij} = \infty$ (or an arbitrary, large number) if the voxels are not in contact.
\item $d_{ij} = 0$ if $i=j$.
\end{itemize}
Once the distance matrix of a graph has been built, it can be used to identify connected subsets of nodes via graph-search algorithms such as \emph{breadth-first search} (BFS) \cite{Cormen:2009}. In our case, these connected components correspond, in general, to the continuous ionization tracks left by charged particles in the active volume of the detector. The BFS algorithm can also be used to calculate the length of the (shortest) path between any pair of voxels in a connected subset. The longest of such paths should connect the ends of a track, that is, the voxels that in a \bbonu-decay track correspond to the blobs. Figure~\ref{fig:Extremes} shows, for simulated electron tracks of energy equal to the $Q$ value of \Xe, the Euclidean distance between the true ends of a track and those found with the BFS algorithm. The most frequent distance in this histogram corresponds to about 6~mm, that is, less than the size of a voxel (1~cm). In other words, in most cases we were able to locate the ends of a track with an accuracy of one or two voxels.

\subsection{Event selection} \label{sec:EventSelection}
The criteria to accept a reconstructed event as a \bbonu-decay candidate are the following:
\begin{enumerate}
\item The event consists of one single reconstructed track confined within the fiducial volume of the detector ---\thinspace defined by excluding a region of 2~cm around the boundaries of the active volume\thinspace--- and with energy between 2.4 and 2.5~MeV.
\item The reconstructed track features a \emph{blob} in both ends.
\item The energy of the event is within the \emph{region of interest} (ROI) around \Qbb.
\end{enumerate}

The definition of a fiducial volume has two purposes: it rejects all charged backgrounds entering the detector and it discards those events in which the tracked particles may have left the active volume, depositing part of their energy in passive materials. The size of the excluded region, 2 centimetres around the boundaries of the active volume, takes into account the voxel size (which, in turn, depends on the spatial resolution of the detector) and the higher inhomogeneity of the electric field near the edges of the field cage (which may affect the quality of the reconstruction in that region). In practical terms, this fiducial cut is implemented demanding that none of the voxels located in the vetoed region contains energy above the detection threshold of the tracking plane (set, conservatively, to 10~keV).

\begin{figure}
\centering
\includegraphics[width=0.5\textwidth]{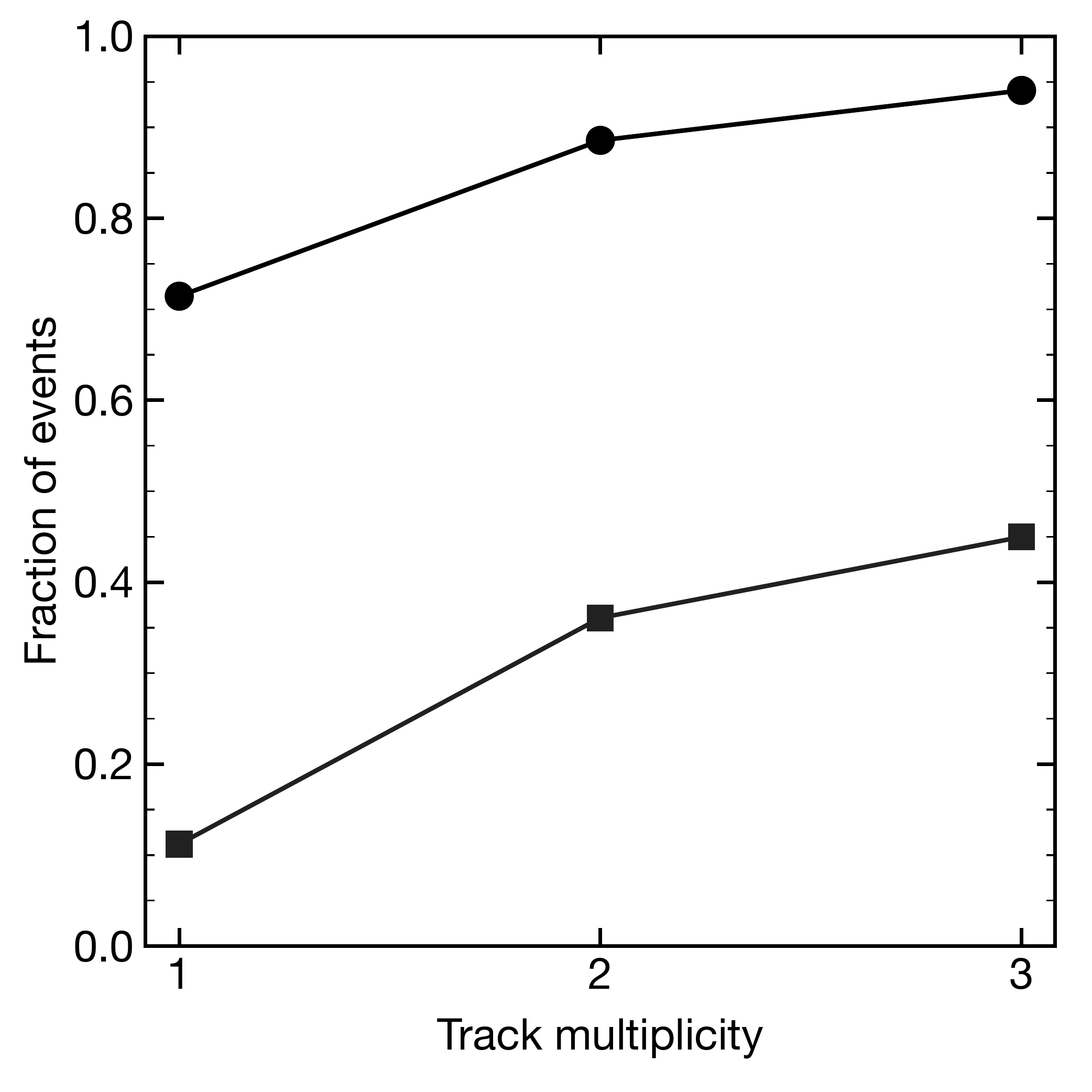}
\caption{Cumulative fraction of signal (dots) and background (squares) events in the fiducial volume as a function of their track multiplicity.} \label{fig:TrackMultiplicity}
\end{figure}

The requirement for the accepted events to have one and only one reconstructed track takes advantage of the very different track multiplicities ---\thinspace that is, the number of reconstructed tracks per event\thinspace--- of signal and background. This is shown in Fig.~\ref{fig:TrackMultiplicity}, where the fraction of events with (at most) a given number of reconstructed tracks is represented for signal and background events contained within the fiducial volume. The latter correspond to \Tl\ and \Bi\ events originating in the PMTs (which, given their location in the detector and contribution to the radioactivity budget, are an important source of background) weighted according to their activities. Approximately 70\% of the signal events satisfy the single-track condition, whereas only 10\% of \Tl\ and \Bi\ events do so. Taking into account that, according to Eq.~(\ref{eq:FigureOfMerit}), maximizing the ratio $\varepsilon_\mathrm{S}/\sqrt{\varepsilon_\mathrm{B}}$ (where $\varepsilon_\mathrm{S}$ and $\varepsilon_\mathrm{B}$ are, respectively, the acceptances of signal and background) optimizes the experimental sensitivity to \mbb, it would only be worth accepting events with more than one track if the fraction of background events passing the cut were such that
\begin{equation}
\varepsilon_\mathrm{B}' \leq (\varepsilon_\mathrm{S}'/\varepsilon_\mathrm{S})^{2}\ \varepsilon_\mathrm{B}\,,
\end{equation}
where the unprimed and primed quantities are, respectively, the acceptances of the default (that is, track multiplicity equal to 1) and the alternative (for instance, track multiplicity equal to 2) selection cuts. For the values shown in Fig.~\ref{fig:TrackMultiplicity} ($\varepsilon_\mathrm{S}=0.71$, $\varepsilon_\mathrm{S}'=0.89$, $\varepsilon_\mathrm{B}=0.11$), we would only improve for $\varepsilon_\mathrm{B}'\leq0.17$. However, the fraction of background events with one or two reconstructed tracks is almost 40\%.

\begin{figure}
\centering
\includegraphics[trim=0 10 0 60, clip, width=0.525\textwidth]{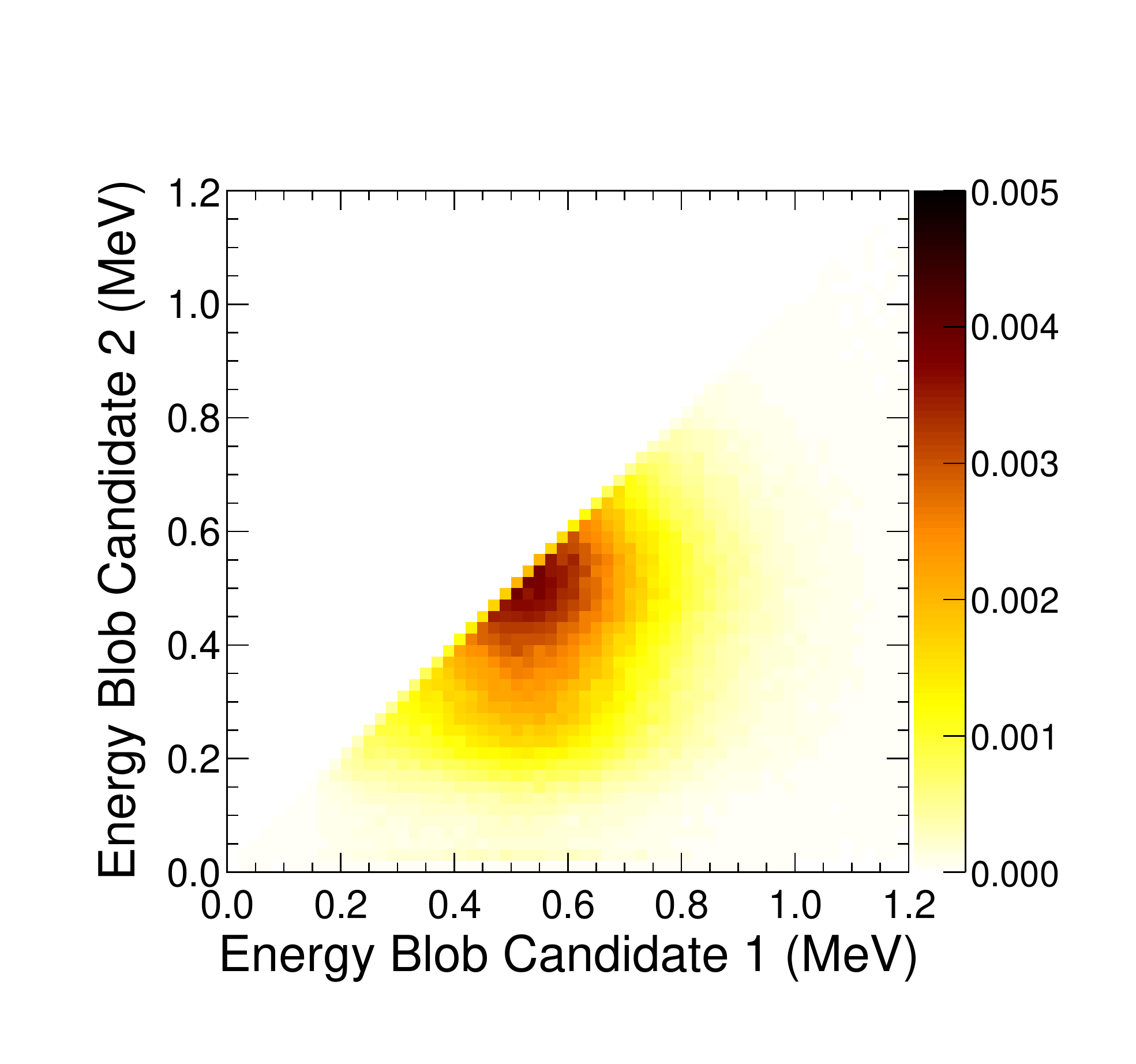}
\includegraphics[trim=0 10 0 60, clip, width=0.525\textwidth]{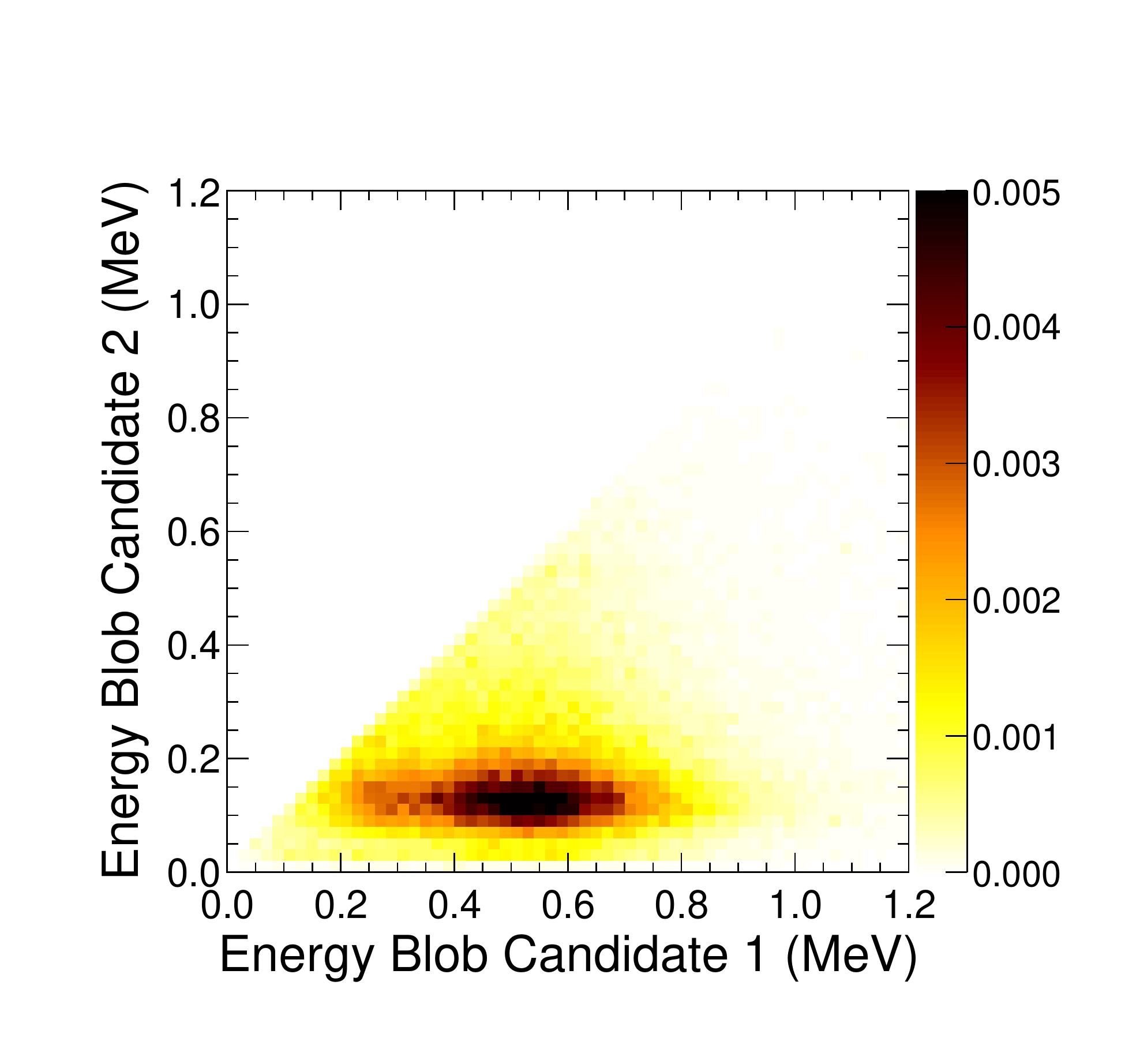}
\includegraphics[trim=0 10 0 60, clip, width=0.525\textwidth]{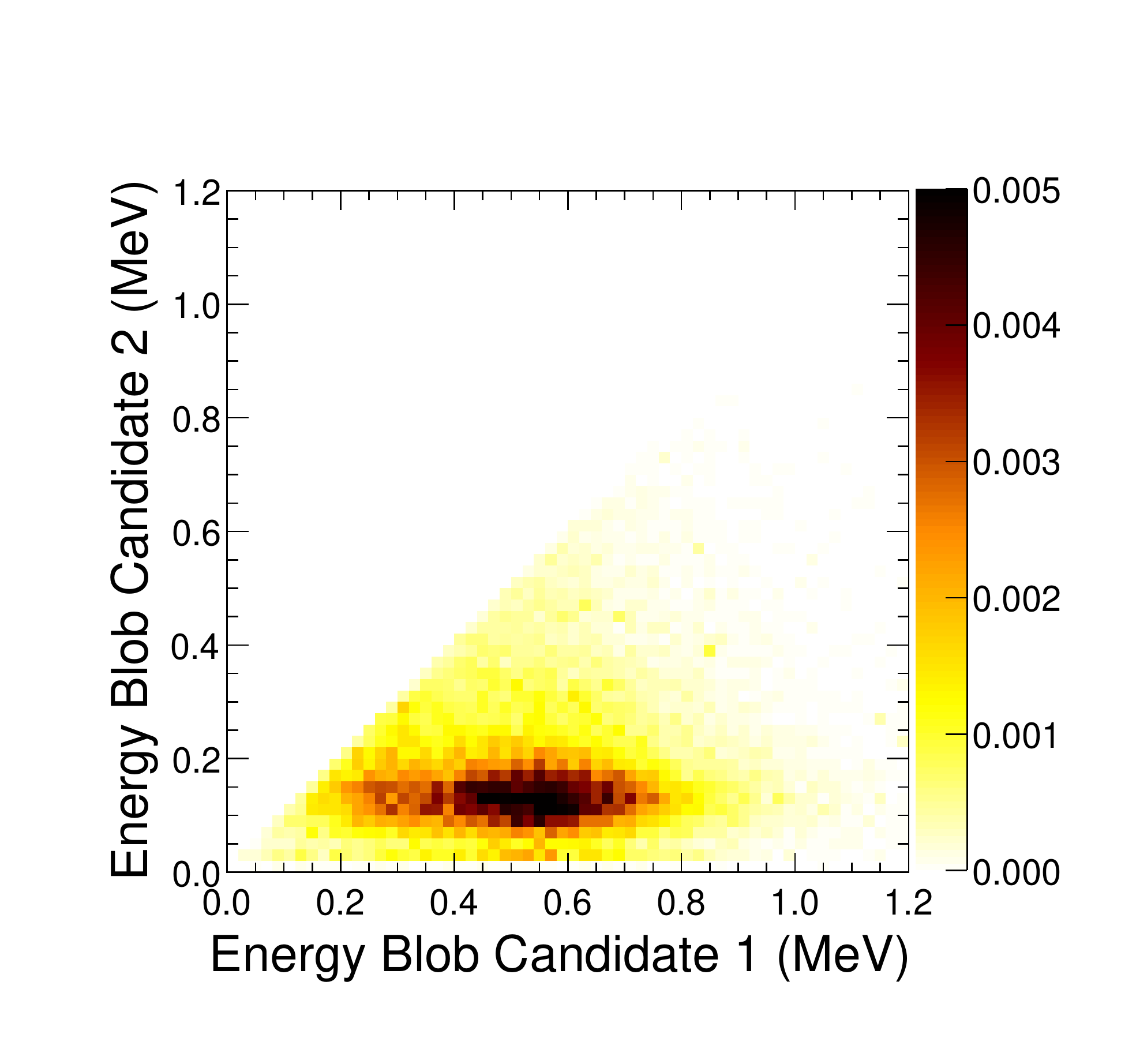}
\caption{Probability distribution of \bbonu\ (top panel), \Tl\ (centre) and \Bi\ (bottom) events in terms of the energies at the end of the tracks. The blob candidate labelled as `1' corresponds to the more energetic one.} \label{fig:EnergyBlobs}
\end{figure}

\begin{figure}
\centering
\includegraphics[width=0.45\textwidth]{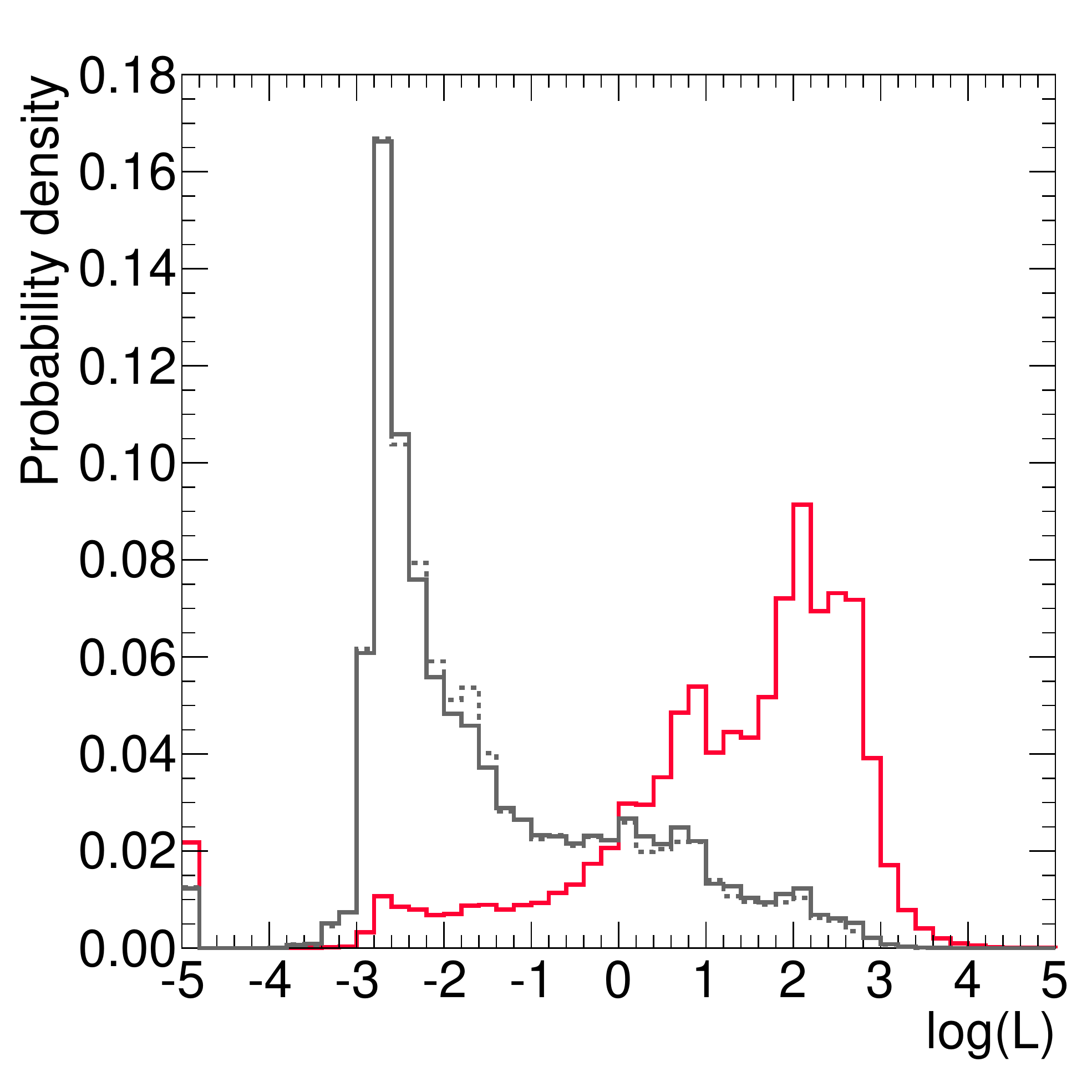}
\caption{Likelihood-ratio distributions for signal (red, solid histogram) and background (\Tl: grey, solid histogram; \Bi: grey, dotted histogram).} \label{fig:LRBlobs}
\end{figure}

The second selection criterion exploits the characteristic energy-deposition pattern of \bbonu-decay tracks (see Fig.~\ref{fig:TrackSignature}), which feature a blob at both ends due to the effective rise in the $\mathrm{d}E/\mathrm{d}x$ of electrons with low momentum. We define the energy of a blob as the total energy contained in all the voxels whose center is at a maximum distance of $\sqrt{3}$~cm with respect to the one reconstructed as track end (that is, all the voxels in contact with it). From the point of view of the discrimination power of the cut, this definition proved to be the optimal among those considered (integration radii between 0 and $\sqrt{3}$~cm). Figure~\ref{fig:EnergyBlobs} shows the probability distributions of signal and background events in terms of the energies of the reconstructed ends of the tracks. The populations of signal and background are clearly separated. Additionally, the distributions of \Tl\ and \Bi\ are very similar, indicating that they correspond to the same type of events (single-electron tracks with energy around \Qbb). A simple and reasonably clean selection cut could be established with a threshold around 0.2~MeV on the energy of the less energetic blob candidate. According to the Neyman-Pearson lemma \cite{Neyman:1933lr}, however, the most efficient selection criterion is based on the likelihood ratio test statistic:
\begin{equation}
\mathcal{L} = \frac{P(E_{1}, E_{2}~|~\bbonu)}{P(\Tl)\cdot P(E_{1},E_{2}~|~\Tl) + P(\Bi) \cdot P(E_{1},E_{2}~|~\Bi)}\,,
\end{equation}
where $P(E_{1}, E_{2}~|~H)$ is the probability for a signal ($H\equiv\bbonu$) or background ($H\equiv{\Tl}$ or $H\equiv{\Bi}$) event to have blob candidates with energies $E_{1}$ and $E_{2}$,  and $P(\Tl)$ and $P(\Bi)$, with $P(\Tl)+P(\Bi)=1$, are the a priori probabilities for a background event to be either \Tl\ or \Bi. In other words, $P(E_{1}, E_{2}~|~H)$ is the probability given in Fig.~\ref{fig:EnergyBlobs}, and $P(\Tl)$ and $P(\Bi)$ are the relative initial abundances of each background source. Once the likelihood ratio ---\thinspace or the natural logarithm of the likelihood function, the so-called log-likelihood, which is, in general, more convenient to work with\thinspace--- is computed for all values of $E_{1}$ and $E_{2}$ (see Fig.~\ref{fig:LRBlobs}), we choose as selection threshold the value of $\mathcal{L}$ that maximizes the figure of merit $\varepsilon/\sqrt{b}$ that is, $\mathcal{L}=1$.

\begin{figure}
\centering
\includegraphics[width=0.65\textwidth]{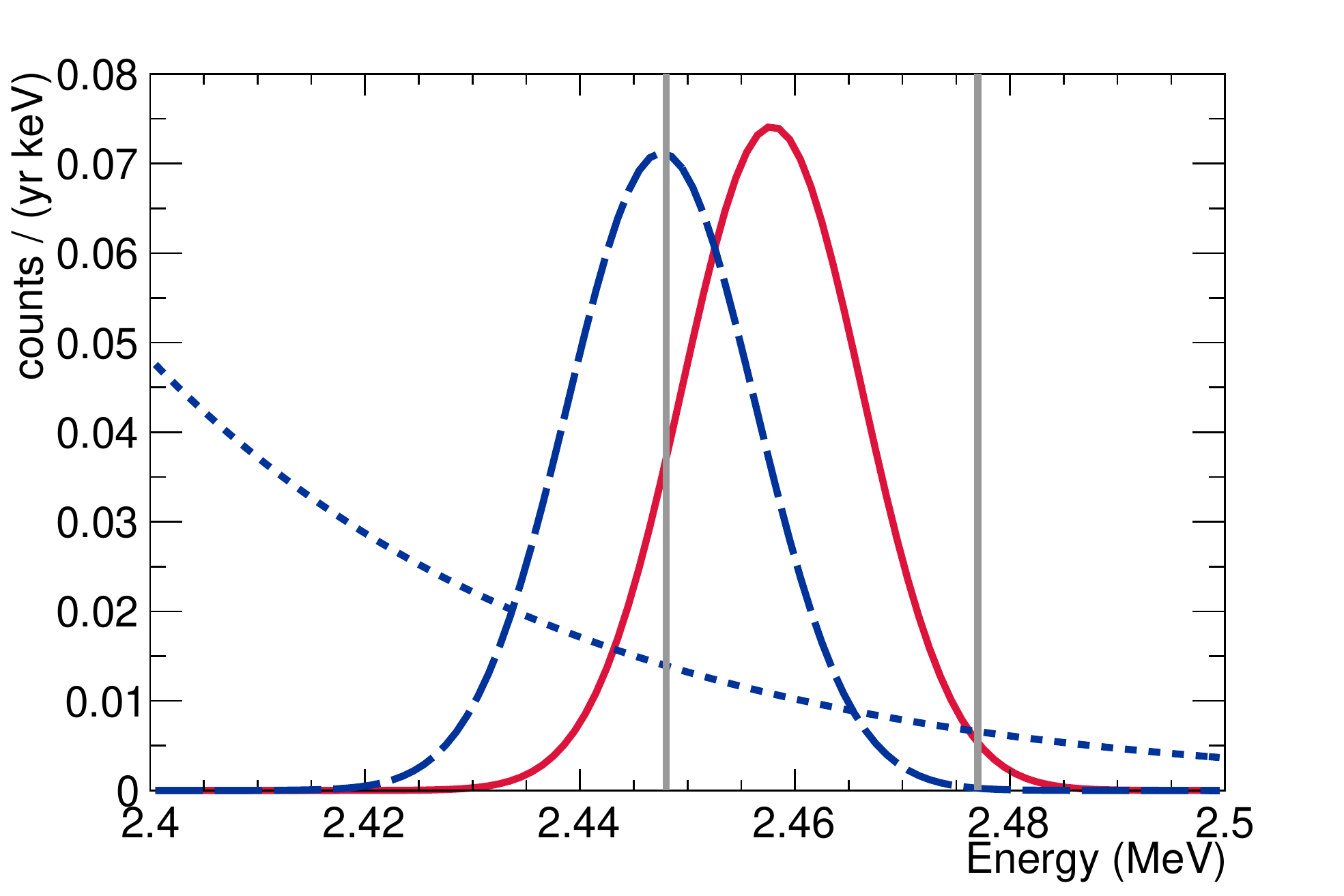}
\caption{Energy distributions of signal (red, solid line) and background (\Tl: blue, dotted line; \Bi: blue, dashed line) in the region around \Qbb\ (2458~keV). The optimal ROI (the one that maximizes the ratio of the signal efficiency over the square root of the background rate) is delimited by the two grey, vertical lines. The signal strength represented here corresponds to a neutrino Majorana mass of 80 meV, while the backgrounds are scaled to their expected values in NEXT-100 ($4\times10^{-4}$~\ckky), assuming an exposure of 91~kg~yr.} \label{fig:ROI}
\end{figure}

A similar procedure can be followed with the third selection criterion in order to decide on the optimal region of interest in the energy spectrum. In this case, the likelihood ratio is defined as follows:
\begin{equation}
\mathcal{L} = \frac{P(E\,|\,\bbonu)}{P(\Tl) \cdot P(E\,|\,\Tl) + P(\Bi) \cdot P(E\,|\,\Bi)}\,,
\end{equation}
where $P(E\,|\,H)$ is the probability for a signal ($H\equiv\bbonu$) or background ($H\equiv{\Tl}$ or $H\equiv{\Bi}$) event to have energy $E$. Figure~\ref{fig:ROI} shows the distribution of signal and background around \Qbb\ and the region of interest that maximizes the quantity $\varepsilon/\sqrt{b}$, selected using the likelihood ratio defined above.

Table~\ref{tab:AcceptanceSelectionCriteria} summarizes the acceptances for signal and background of the selection criteria described above. The natural radioactive backgrounds, \Tl\ and \Bi, are suppressed by more than 6 orders of magnitude, and the contribution of \bbtnu-decay to the background rate is completely negligible. The cuts yield a signal efficiency of 28\%. Note, however, that approximately half of the events are lost already in the first selection cut: 88\% of the events are contained within the fiducial volume of the detector, 71\% have one single track and 76\% of them have reconstructed energy above 2.4~MeV (the \bbonu\ spectrum has a tail extending to low energies composed of events with missing energy in the form of bremsstrahlung radiation). More sophisticated analyses ---\thinspace making use of multivariate statistics, for instance\thinspace--- can probably improve somewhat the signal selection efficiency and background rejection power obtained here. This study, however, is still affected by large uncertainties (in particular, those arising from radio-screening measurements) and, in addition, the simple approach presented in this section has the advantage of revealing the individual power of the energy and tracking event signatures.

\begin{table}
\centering
\begin{tabular}{l c l l l}
\toprule
Selection criterion & \multicolumn{1}{c}{\bbonu} & \multicolumn{1}{c}{\bbtnu} & \multicolumn{1}{c}{\Tl} & \multicolumn{1}{c}{\Bi} \\ \midrule
Fiducial, single track & \multirow{2}{*}{0.4759} & \multirow{2}{*}{$8.06\times10^{-9}$} & \multirow{2}{*}{$1.39\times10^{-5}$} & \multirow{2}{*}{$3.40\times10^{-6}$} \\
$E\in[2.4, 2.5]~\mathrm{MeV}$ \\ \addlinespace
Track with 2 blobs & 0.6851 & 0.6851 & 0.1141 & 0.1005 \\ \addlinespace
Energy ROI & 0.8661 & $3.89\times10^{-5}$ & 0.1515 & 0.4795 \\ \addlinespace
\emph{Total} & 0.2824 & $2.15\times10^{-13}$ & $2.4\times10^{-7}$ & $1.6\times10^{-7}$ \\
\bottomrule
\end{tabular}
\caption{Acceptance of the selection criteria for \bbonu-decay events described in the text. The figures for \Tl\ and \Bi\ correspond to background events originating in the PMTs, one of the dominant sources of background in the detector.} \label{tab:AcceptanceSelectionCriteria}
\end{table}

\section{Estimated background rate and sensitivity to \mbb} \label{sec:BkgRate}
\begin{figure}
\centering
\includegraphics[width=0.825\textwidth]{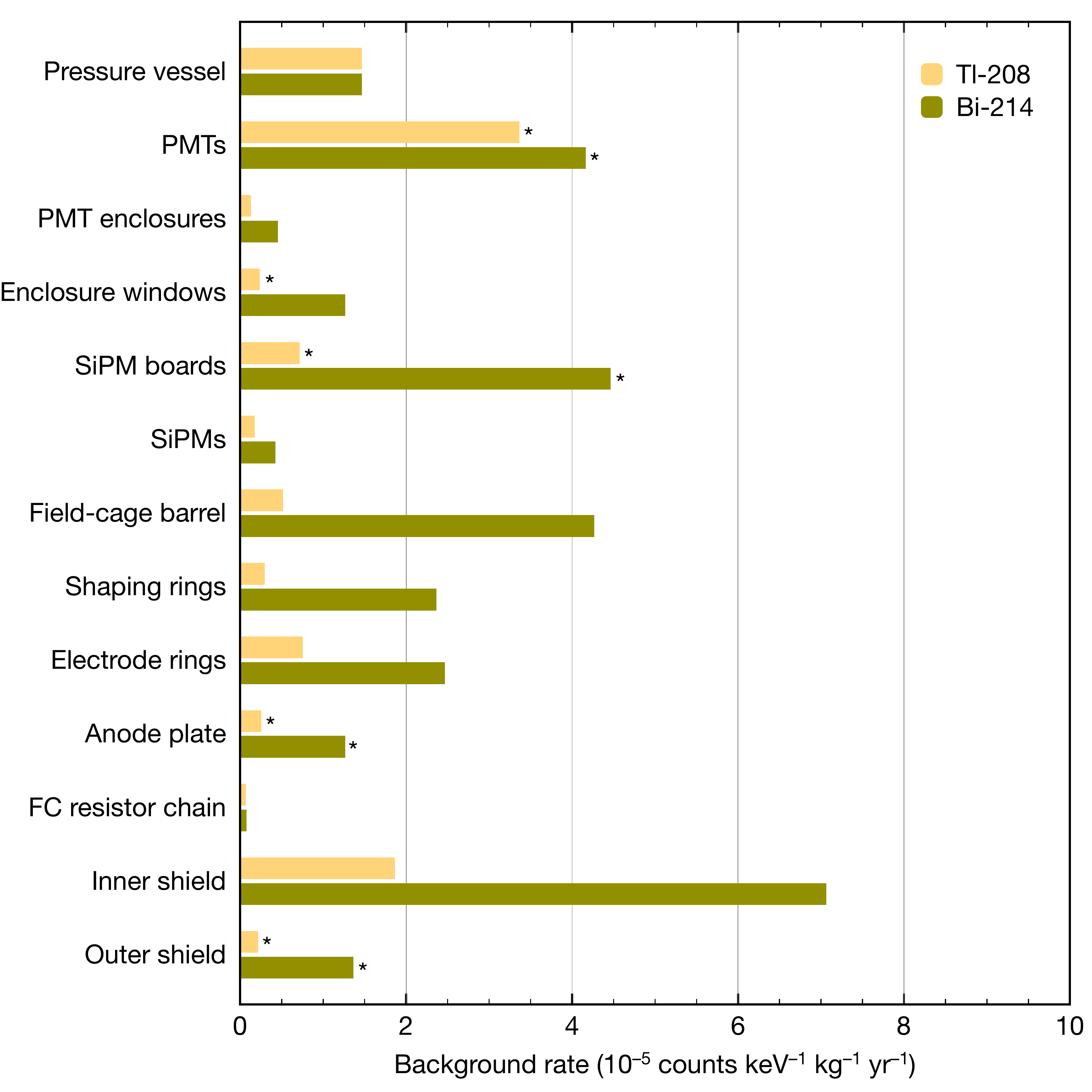}
\caption{Contribution to the background rate of NEXT-100 of the different detector subsystems considered in our background model. An asterisk (*) next to a bar indicates that the contribution corresponds to a positive measurement of the activity of the material.} \label{fig:BackgroundContributions}
\end{figure}

The contribution of each detector subsystem to the overall background rate of NEXT-100 is shown in Table~\ref{tab:BackgroundContributions}. These rates are obtained dividing the initial activities of \Tl\ and \Bi\ by the corresponding background rejection factors (defined as the inverse of the background acceptance resulting from the \bbonu-decay event selection described in the previous section). They are also represented graphically in Fig.~\ref{fig:BackgroundContributions}. Notice that our knowledge is quite uncertain, given that for many background sources we only have at present a limit to their activity. This is, in fact, a problem common to all \bbonu-decay experiments, and it will be even more serious for the experiments of the tonne scale, which will require materials and components of higher radiopurity.

\begin{table}
\centering
\footnotesize
\begin{tabular*}{\textwidth}{@{\extracolsep{\fill}} l *{2}{c} *{2}{c} *{2}{D{.}{.}{4.4}}}
\toprule
Detector subsystem & \multicolumn{2}{c}{Activity (mBq)} & \multicolumn{2}{c}{Rejection factor ($10^6$)} & \multicolumn{2}{c}{$c$ $\left(\mathrm{10^{-4}/(keV~kg~yr)}\right)$} \\ \cmidrule(lr){2-3} \cmidrule(lr){4-5} \cmidrule(l){6-7}
       & \multicolumn{1}{c}{\Tl} & \multicolumn{1}{c}{\Bi} & \multicolumn{1}{c}{\Tl} & \multicolumn{1}{c}{\Bi} & \multicolumn{1}{c}{\Tl} & \multicolumn{1}{c}{\Bi} \\ \midrule
\emph{Pressure vessel} \\
\quad Total & $<197$ & $<603$ & $170(80)$ & $500(400)$ & <0.14 & <0.14 \\ \addlinespace
\emph{Energy plane} \\
\quad PMTs & $18(5)$ & $<56$ & $4.1(3)$ & $6.1(5)$ & 0.33(8) & 0.41(10) \\
\quad PMT enclosures & $<0.36$ & $<3.1$ & $6.8(6)$ & $9.5(9)$ & <0.006 & <0.04 \\
\quad Enclosure windows & $0.34(8)$ & $<2.6$ & $2.38(11)$ & $2.56(13)$ & 0.017(4) & <0.12 \\ \addlinespace
%
\emph{Tracking plane} \\
\quad SiPMs & $<0.2$ & $<0.6$ & $2.04(8)$ & $2.04(8)$ & <0.011 & <0.036 \\
\quad SiPM boards & $1.11(12)$ & $7.5(5)$ & $2.04(8)$ & $2.04(8)$ & 0.065(7) & 0.44(4) \\ \addlinespace
\emph{Electric-field cage} \\
\quad Barrel & $<1$ & $<8$ & $2.61(13)$ & $2.27(10)$ & <0.05 & <0.4 \\ 
\quad Shaping rings & $<0.5$ & $<4$ & $2.61(13)$ & $2.27(10)$ & <0.023 & <0.23 \\
\quad Electrode rings & $<1.5$ & $<5$ & $2.61(13)$ & $2.27(10)$ & <0.07 & < 0.24 \\
\quad Anode plate & $0.32(4)$ & $2.0(5)$ & $2.04(8)$ & $2.04(8)$ & 0.019(2) & 0.12(3) \\
\quad Resistor chain & $<0.0026$ & $<0.02$ & $2.61(13)$ & $2.27(10)$ & <0.00012 & <0.0011 \\ \addlinespace
\emph{Shielding} \\
\quad Inner shield & $<13$ & $<120$ & $8.9(8)$ & $19(3)$ & <0.18 & <0.7 \\
\quad Outer shield & $2060(430)$ & $21300(4300)$ & $17000(5000)$ & $20000(5000)$ & 0.015(3) & 0.13(3) \\
\bottomrule
\end{tabular*}
\normalsize
\caption{Contribution to the background rate of NEXT-100 predicted for each subsystem of the detector considered in our background model. The second and third columns correspond to the initial activities of \Tl\ and \Bi\ (see Table~\ref{tab:RadioactivityBudget}). The fourth and fifth columns contain the rejection factors computed with the detector simulation; the figures in parenthesis correspond to the statistical error. The last two columns in the table show the background rate estimated for each subsystem (i.e.\ the ratio of the previous quantities) expressed in $10^{-4}$~\ckky. For most subsystems, we only have upper limits to their induced background rate. In those cases where we have a positive measurement, the figures in parentheses give the 1-sigma uncertainty in the last digit.} \label{tab:BackgroundContributions}
\end{table}

Table~\ref{tab:BkgRateSummary} shows the contributions grouped into six major subsystems. The background from \Bi\ is about three times larger than the background from \Tl. The overall background rate estimated for NEXT-100 is
\begin{equation}
<4\times10^{-4}~\mathrm{counts/(keV~kg~year)}\,.
\end{equation}
The above rate includes only radioactive backgrounds from detector materials and components. All other sources of background are expected to contribute each one at the level of $10^{-5}$~keV$^{-1}$~kg$^{-1}$~yr$^{-1}$ or below.

\begin{table}
\centering
\begin{tabular*}{.8\textwidth}{@{\extracolsep{\fill}} l *{3}{D{.}{.}{4.6}}}
\toprule
Detector subsystem & \multicolumn{1}{c}{\Tl} & \multicolumn{1}{c}{\Bi} & \multicolumn{1}{c}{\itshape Total} \\ \midrule
Pressure vessel     & <0.14 & <0.14 & <0.28 \\
Energy plane        & <0.37 & <0.61 & <0.98 \\
Tracking plane      & <0.08 & <0.48 & <0.56 \\
Electric-field cage & <0.16 & <1.00 & <1.16 \\
Inner shield        & <0.18 & <0.73 & <0.91 \\
Outer shield        & 0.015(3) & 0.130(30) & 0.140(30) \\
{\itshape Total}    & <0.94 & <3.09 & <4.03 \\ 
\bottomrule
\end{tabular*}
\caption{Estimated contribution of major detector subsystems to the background rate of NEXT-100, expressed in 10$^{-4}$ counts~keV$^{-1}$~kg$^{-1}$~yr$^{-1}$.} \label{tab:BkgRateSummary}
\end{table}

\begin{figure}
\centering
\includegraphics[width=0.8\textwidth]{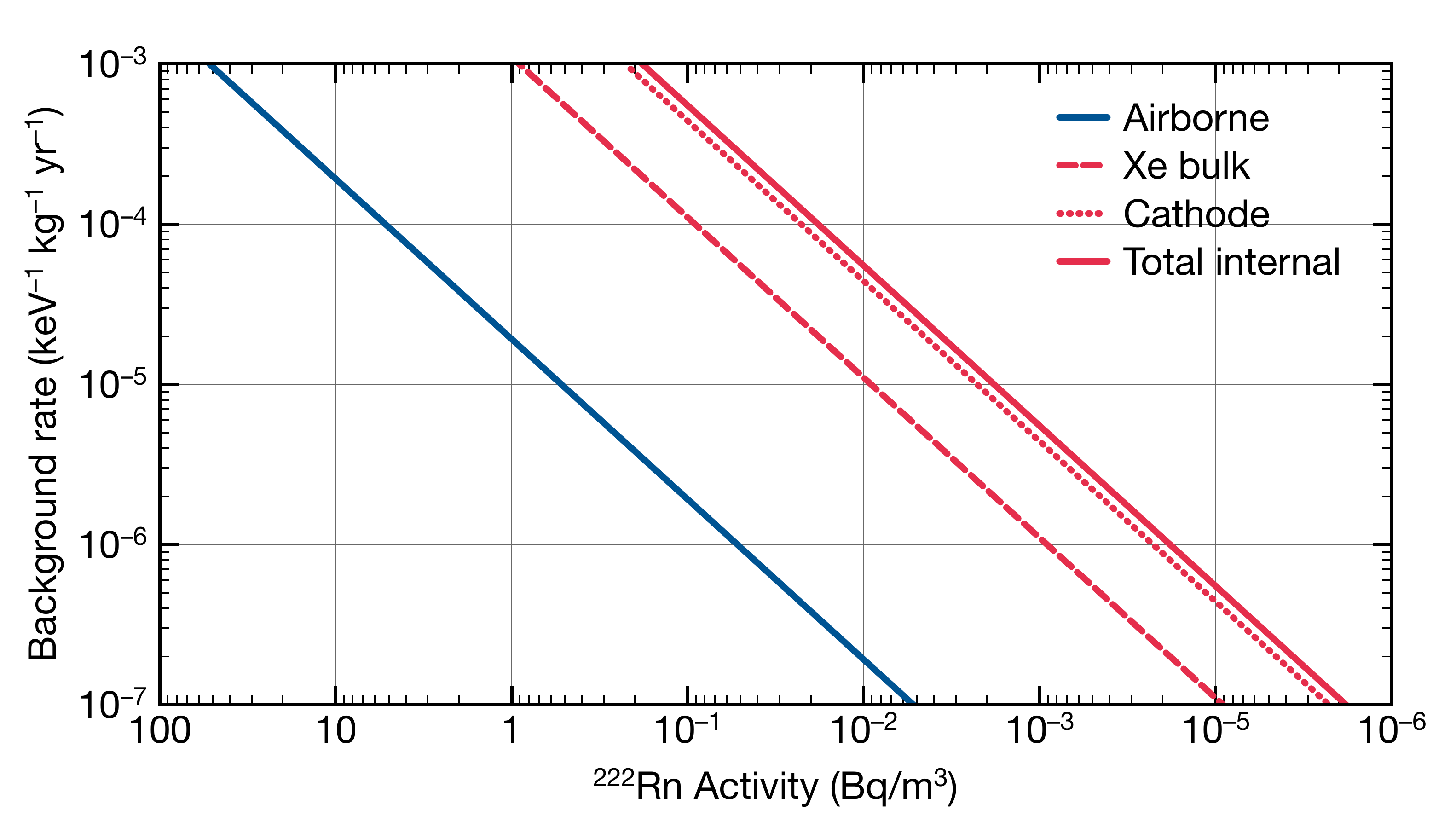}
\caption{Background rate induced in NEXT-100 by airborne radon and radon contamination in the xenon gas (labelled as \emph{internal}) in terms of the activity of $^{222}$Rn.} \label{fig:RadonNext100}
\end{figure}

In particular, the activity of airborne radon in the vicinity of the detector ---\thinspace which translates, ultimately, into \Bi\ activity on the internal surface of the lead shield and on the external surface of the vessel\thinspace--- will be reduced by at least two orders of magnitude with respect to the activity in the experimental hall of LSC ($\sim$\thinspace80~Bq/m$^{3}$) thanks to the use of a radon abatement machine that will be installed in 2016. The computed rejection factor for this source of background is $2\times10^{9}$, resulting in a background rate of about $10^{-5}$~keV$^{-1}$~kg$^{-1}$~yr$^{-1}$ for a $^{222}$Rn activity of 0.5~Bq/m$^{3}$ (see Fig.~\ref{fig:RadonNext100} for other values of the specific activity of $^{222}$Rn in the range between 10$^{-2}$ and 10$^{2}$~Bq/m$^{3}$). Radon contamination in the xenon gas causes two different types of background events: $\beta$ tracks from the decay of \Bi\ in the active volume, and photoelectrons generated by gamma rays emitted, for the most part, from the TPC cathode following the decay of \Bi. In the EXO-200 TPC, the latter type of events constitute about 80\% of the measured activity of $^{222}$Rn in the liquid xenon, while the former make up the remaining 20\% \cite{Albert:2013gpz}. The rejection power against both types of background events is similar, approximately $2.5\times10^{6}$. In the case of the $\beta$ decays of \Bi\ in the xenon bulk, we have estimated that Bi-Po tagging ---\thinspace i.e.\ the coincident detection in an event of the $\beta$ emitted in the decay of \Bi\ and the alpha emitted by $^{214}$Po shortly after\thinspace--- can be done with high efficiency ($\gtrsim99\%$). Figure~\ref{fig:RadonNext100} (red lines) shows the background rate generated in NEXT-100 by this internal contamination of radon in terms of the activity of $^{222}$Rn. In order for this background to contribute, at most, at the level of $10^{-5}$~keV$^{-1}$~kg$^{-1}$~yr$^{-1}$, radon activities in the xenon gas below a few mBq per cubic metre will be required. The EXO-200 detector, which has been operating without a radon suppression system, has measured, for instance, an activity of $^{222}$Rn of that order in their xenon volume: $(3.65\pm0.37)~\mu\mathrm{Bq/kg}$ \cite{Albert:2013gpz}. Similarly, the radon activity of the NEMO-3 tracking gas was measured to be about 5~mBq/m$^{3}$ \cite{Arnold:2013dha}.

\begin{figure}
\centering
\includegraphics[width=0.525\textwidth]{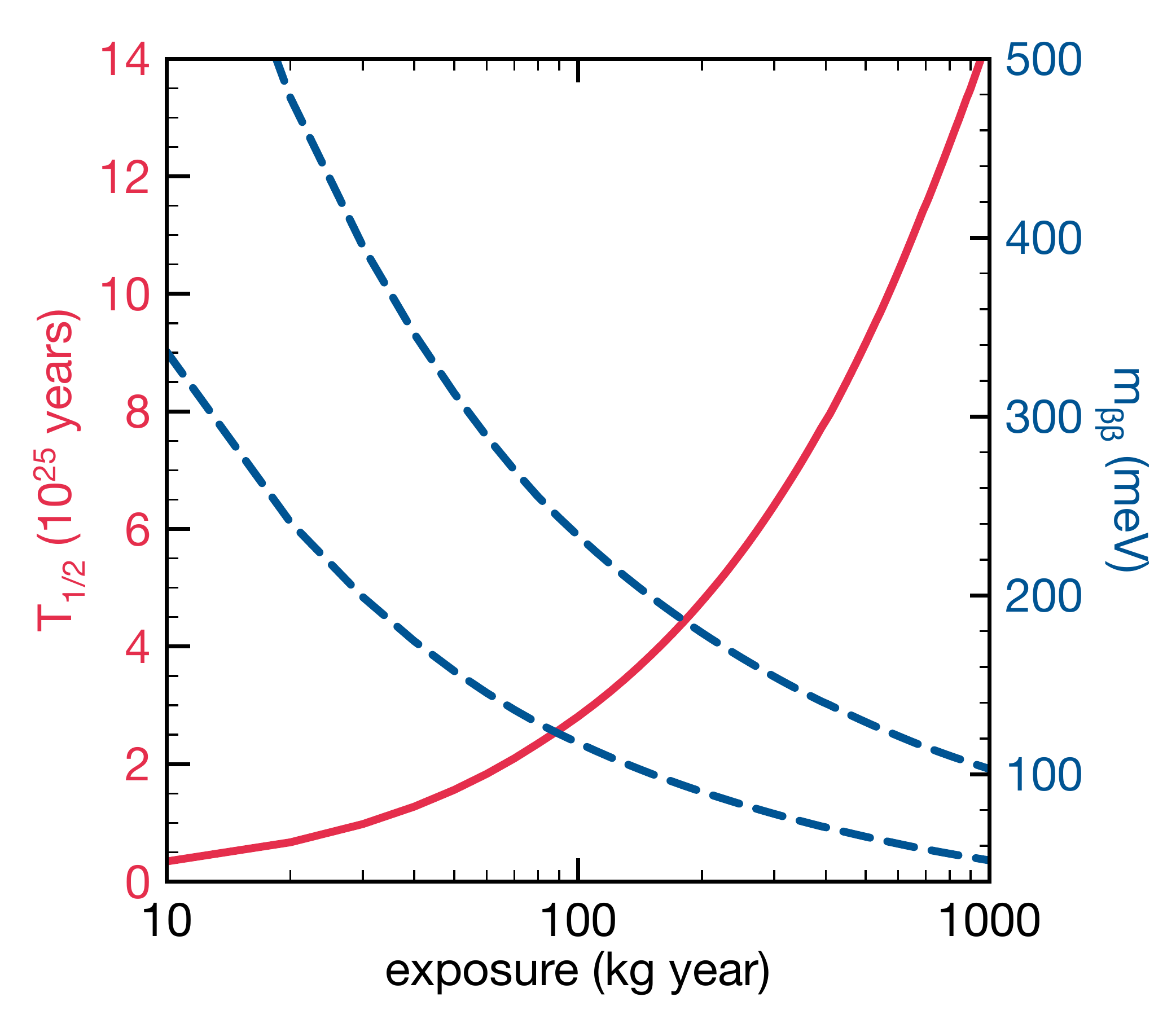}
\includegraphics[width=0.525\textwidth]{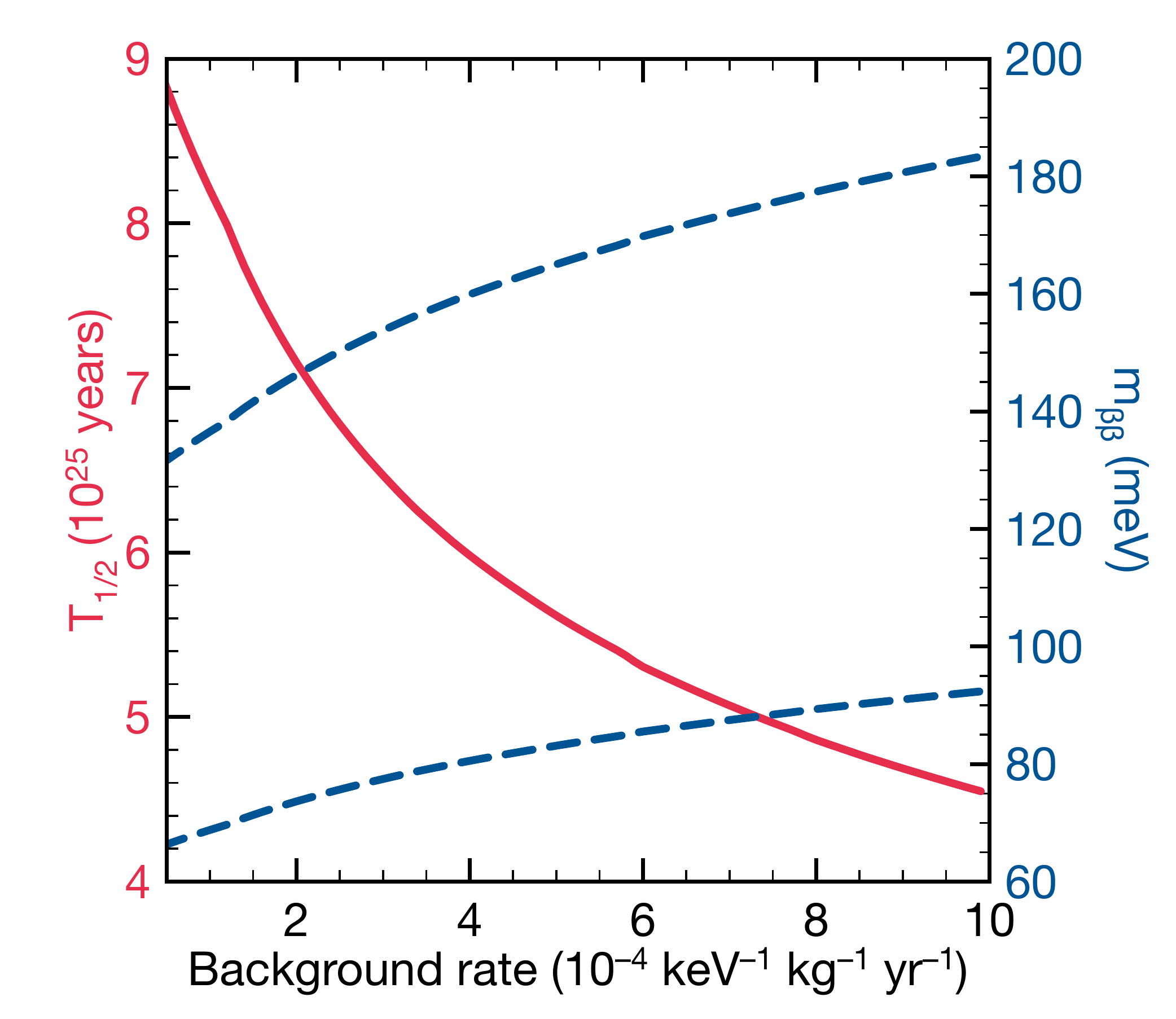}
\caption{Sensitivity (at 90\% CL) of NEXT-100 to neutrinoless double beta decay.  The (crimson) solid curves represent the half-life sensitivity, while the (blue) dashed curves correspond to the \mbb\ sensitivity for the largest (EDF) and smallest (ISM) NME estimates listed in Table~\ref{tab:Xe136}. Top: Sensitivity of NEXT-100 in terms of the accumulated exposure for an estimated background rate of $4\times10^{-4}$~\ckky. Bottom: Sensitivity after an effective 3-year run (equivalent to an exposure of about 275~kg~yr) as a function of the achieved background rate.} \label{fig:SensitivityNEXT100}
\end{figure}

The sensitivity of NEXT-100 to neutrinoless double beta decay ---\thinspace calculated following the Feldman-Cousins prescription for the construction of confidence intervals \cite{Feldman:1997qc, GomezCadenas:2010gs}\thinspace--- is shown in Fig.~\ref{fig:SensitivityNEXT100}. In the top panel, the sensitivity (at 90\% CL) to the half-life (red, solid curve) and the corresponding sensitivity to \mbb\ (blue, dashed curves) for the largest and smallest NME calculations in Table~\ref{tab:Xe136} are represented in terms of the exposure, assuming a signal detection efficiency of 28\% and a background rate of $4\times10^{-4}~\ckky$. Since the uncertainties associated to material screening measurements are large, it could well be that the actual background rate due to natural radioactivity is noticeably smaller than the upper limit derived in this paper once the detector is operated (and that other sources of background become then relevant). For this reason, the bottom panel in Fig.~\ref{fig:SensitivityNEXT100} shows, for an exposure of 275~kg~yr, the variation of the sensitivity with respect to the achieved background rate in the range between $10^{-4}$ and $10^{-3}$~\ckky. The NEW data will provide in the near future a quantitative assessment of this question.

\section{Summary and conclusions} \label{sec:Conclusions}
The importance of the worldwide experimental program searching for neutrinoless double beta decay can hardly be overstated: the discovery of the radioactive process is the most promising way ---\thinspace perhaps the only way\thinspace--- of establishing the nature of neutrino mass and determining whether total lepton number is violated. In this paper, we have discussed in depth one of the new-generation experiments: the Neutrino Experiment with a Xenon TPC (NEXT), which will search for the neutrinoless double beta decay of \Xe\ at the Laboratorio Subterr\'aneo de Canfranc (LSC). NEXT possesses two features of great value for \bbonu-decay searches: excellent energy resolution and an extra experimental signature, charged-particle tracking, for the active suppression of background. The goal of the NEXT project is the construction, commissioning and operation of the NEXT-100 detector, a 100-kg xenon gas TPC built with radiopure materials that will start taking low-background data at LSC in 2018. Prior to that, the NEXT Collaboration will operate the NEXT-White (NEW) detector, a technology demonstrator that implements in a 1:2 scale the design chosen for NEXT-100 using the same materials and photosensors. The NEW data will make possible the validation of the NEXT-100 background model, currently based on detailed Monte Carlo detector simulation and material-screening measurements that predict a background rate for NEXT-100 of at most $4\times10^{-4}$~\ckky. With this background rate, the NEXT-100 detector will reach a sensitivity (at 90\% CL) to the \bbonu-decay half-life of $2.8\times10^{25}$~years for an exposure of 100 $\mathrm{kg}\cdot\mathrm{year}$, or $6.0\times10^{25}$~years after running for 3 effective years. This corresponds to an upper limit on the Majorana neutrino mass of 80--160 meV, depending on the used NME calculation. We believe nonetheless that there is ample room for improvement with respect to the baseline detector performance described in this work. With the use of more sophisticated reconstruction and selection algorithms, currently under development, it should be possible to reach an energy resolution close to 0.5\% FWHM at 2.5 MeV and fully exploit the potential of the tracking signature. 


\acknowledgments
The NEXT Collaboration acknowledges support from the following agencies and institutions: the European Research Council (ERC) under the Advanced Grant 339787-NEXT; the Ministerio de Econom\'ia y Competitividad of Spain under grants CONSOLIDER-Ingenio 2010 CSD2008-0037 (CUP), FIS2014-53371-C04 and the Severo Ochoa Program SEV-2014-0398; the Portuguese FCT and FEDER through the program COMPETE, project PTDC/FIS/103860/2008; the U.S.\ Department of Energy under contracts number DE-AC02-07CH11359 (Fermi National Accelerator Laboratory) and DE-FG02-13ER42020 (Texas A\&M); and the University of Texas at Arlington. We thank Joshua B.\ Albert for useful discussions regarding the cosmogenic backgrounds.


\bibliography{refs}
\bibliographystyle{JHEP}

\end{document}